\documentclass[amsmath,amssymb,amsfonts, twocolumn, superscriptaddress]{revtex4}
\usepackage[german,american,english]{babel}
\usepackage{graphicx}
\usepackage{graphics}
\usepackage{dcolumn}
\usepackage{bm}
\usepackage{amssymb}
\usepackage{amsmath}
\usepackage{amsfonts}
\usepackage{epsfig}
\usepackage{color}

\newcommand {\bra} [1] {\langle #1 |}
\newcommand {\ket} [1] {| #1 \rangle}
\newcommand {\bkt} [1] {\langle #1 \rangle}

\newcommand {\tbkt} [3] {\langle #1 | #2 | #3 \rangle}

 \newcommand {\beq}{\begin{equation}}
\newcommand {\eeq}{\end{equation}}

\begin{document}
\title{Quantum Computing with Acceptor Spins in Silicon}

\author{Joe Salfi}
\affiliation{School of Physics, The University of New South Wales, Sydney 2052, Australia}

\author{Mengyang Tong}
\affiliation{School of Physics, The University of New South Wales, Sydney 2052, Australia}
\affiliation{Department of Physics, Beijing Normal University, Beijing, 100875, P. R. China}

\author{Sven Rogge}
\affiliation{School of Physics, The University of New South Wales, Sydney 2052, Australia}

\author{Dimitrie Culcer}
\affiliation{School of Physics, The University of New South Wales, Sydney 2052, Australia}

\begin{abstract}
The states of a boron acceptor near a Si/SiO$_2$ interface, which bind two low-energy Kramers pairs, have exceptional properties for encoding quantum information and, with the aid of strain, both heavy hole and light hole-based spin qubits can be designed. Whereas a light-hole spin qubit was introduced recently [Phys. Rev. Lett. \textbf{116}, 246801 (2016)], here we present analytical and numerical results proving that a heavy-hole spin qubit can be reliably initialised, rotated and entangled by electrical means alone. This is due to strong Rashba-like spin-orbit interaction terms enabled by the interface inversion asymmetry. Single qubit rotations rely on electric-dipole spin resonance (EDSR), which is strongly enhanced by interface-induced spin-orbit terms. Entanglement can be accomplished by Coulomb exchange, coupling to a resonator, or spin-orbit induced dipole-dipole interactions. By analysing the qubit sensitivity to charge noise, we demonstrate that interface-induced spin-orbit terms are responsible for sweet spots in the dephasing time $T_2^*$ as a function of the top gate electric field, which are close to maxima in the EDSR strength, where the EDSR gate has high fidelity. We show that both qubits can be described using the same starting Hamiltonian, and by comparing their properties we show that the complex interplay of bulk and interface-induced spin-orbit terms allows a high degree of electrical control and makes acceptors potential candidates for scalable quantum computation in Si. 
\end{abstract}

\maketitle

\section{Introduction}
Quantum computing is among the most intensively researched topics in modern physics.\cite{Nielsen_Chuang} The need for scalability and long coherence times has spurred the development of spin-based solid state quantum bits (qubits), \cite{Kane_Nature98, Loss_PRA98, Petta_Science05, Koppens_PRL08, Awsch_Qbt_Rvw_Sci13,Shulman:2012fk,Waldherr:2014kt,Veldhorst:2015jea} which are inherently scalable and interact weakly with their environment. Solid state spin qubits employ the nuclear spin of a donor, \cite{Kane_Nature98} the spin of a quantum dot or donor-bound electron, \cite{Loss_PRA98} or two-electron singlet and triplet states. \cite{Petta_Science05} Among semiconducting host materials, Si is promising due to its compatibility with Si microelectronics, the absence of piezoelectric coupling to phonons and nuclear-spin free isotopes offering the possibility of isotopic purification to eliminate the hyperfine interaction leaving nearly pure $^{28}$Si. \cite{Itoh_MRS} The outstanding spin coherence times of Si \cite{Feher_PR59, Abe_PRB04, Tahan_PRB05, Tyryshkin_JPC06, Wang_SiQD_ST_Relax_PRB10, Raith_SiQD_1e_SpinRelax_PRB11, Wilamowski_Si/SiGeQW_Rashba_PRB02, Witzel_AHF_PRB07, Muhonen_Store_14,Veldhorst:2014eq, Kha_APL15} have caused Si-based devices to be energetically investigated. \cite{Morton_Si_QC_QmLim_Nat11, Zwanenburg_SiQmEl_RMP13, Muhonen_Store_14,Veldhorst:2014eq,Hao_SiDQD_SpinVlly_NC14,Golding:2003tq, Ruskov:2013kq}

At the same time, the study of localised spin systems with strong spin-orbit interactions has flourished in recent years, motivated by their potential for all-electrical spin manipulation. \cite{Kyrychenko_PRB04, Friesen_Spin_Readout_PRL04, Nadj_Nat10, Palyi_SO_Res_PRL12, Bulaev_PRL07, Szumniak_PRL12, Budich_PRB12,Tang:2006hi} Spin-based quantum computing schemes typically rely on static magnetic fields to achieve a splitting of the qubit states, and on radio-frequency (RF) magnetic fields to effect transitions between these states. Static magnetic fields can be localised by using nanomagnets \cite{Pioro_NatPhys08} but localising RF magnetic fields is considerably harder. Hence, a significant body of research has been devoted to accomplishing spin manipulation entirely by electrical means, such as electric dipole spin resonance (EDSR) \cite{Bulaev_PRL07,Nowack:2007du,Medford:2013eka,Pribiag:2013if, Romhanyi_EDSR_15}. Many additional advantages of spin-orbit coupling have been identified. Via coupling to electric fields and deformation potentials, spin orbit coupling opens new channels for entanglement of spin qubits by direct Coulomb interactions, photons, and phonons. When intrinsic spin-orbit coupling of the crystal is employed, complex fabrication of nano-magnets can be avoided, while also avoiding additional decoherence channels they may introduce.

Confined hole systems are ideal to explore spin-orbit coupling effects in localized spins. Hole spin-orbit interactions are inherently strong, due to the orbital angular momentum $l=1$ of their constituent atomic wave functions, and holes are described by an effective spin-3/2 \cite{Winkler2003}. The additional degrees of freedom of the spin-3/2 -- the quadrupole moment associated with the heavy hole-light hole coupling as well as the octupole moment -- give rise to physics unique to holes: an alternating spin polarisation \cite{Culcer_AltSpinPol_PRL06} and interactions with an electric field quadratic in the effective spin.\cite{Bir_JPCS63_2, Winkler2003} Decoherence in hole systems can be less pronounced than in some electron systems, \cite{Tartakovskii_Hole_NPt11, Wang_HH_Echo_PRL12, Chekhovich_NM13} due to the suppression of the hole-nuclear spin contact interaction. \cite{Coish_Nuclear_PSSB09,Tyryshkin:2011fi,Chekhovich:2012ks} Thanks to these favourable properties, hole quantum dots and acceptors have been the focus of increasing attention, \cite{Spruijtenburg_APL13, Pribiag:2013if, MacLeod_APL14, Ruskov:2013kq, Miserev_g_15, Calderon_arXiv15, Voisin_Hole_MOS_15} for example experiments have successfully measured the Zeeman splitting of a single acceptor hole \cite{vanderHeijden:2014fp}.

\begin{figure}
\includegraphics[scale=.8, width = \columnwidth
]{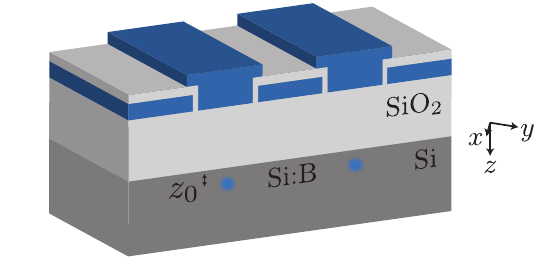}
\caption{Sketch of the qubit architecture, projected onto the $xz$-plane, with $\hat{z}$ perpendicular to the interface.}
\label{Device}
\end{figure}

Spin-orbit coupling in conventional spin qubits enhances unwanted coupling to electric fluctuations arising from phonons and charge fluctuations, thus potentially destroying single- and two-qubit coherence. The fundamental question to be answered is whether strong spin-orbit must necessarily be associated with dephasing and the loss of quantum information,\cite{Culcer_SpinPolDecay_PRB07, Huang_PRB2014, Peihao_SpinValley_PRB14, Bermeister_APL14} or whether quantum computation schemes could be devised in which electrical spin control by desired electric fields is enabled while dephasing due to unwanted electric fields is suppressed. The long-term aim of our work is therefore to devise a highly coherent localised spin-orbit qubit which can be operated and entangled purely through the use of electric fields. In particular it is important that the qubit be robust to electrical noise during both single qubit and two-qubit gates. Protection during two qubit operations could prove to be very important since the exchange mechanism for entangling spin qubits is inherently vulnerable to dephasing by electric field fluctuations.

In this paper we propose using holes bound to B:Si acceptors implanted near an interface (Fig.~\ref{Device}) to realize highly coherent qubits that utilise both the spin and the charge degrees of freedom for enhanced functionality \cite{Morello:2010ga,Fuechsle:2012bl, Pavlov:2014ju,Pla:2012jj} We discuss analytically the various hole interactions with the interface and electric fields and introduce their use in a new heavy-hole qubit, in a manner broadly analogous to our previously proposed light-hole qubit. \cite{Salfi_LH_2015} We develop a theoretical picture unifying acceptor qubits and compare the heavy-hole and light-hole architectures in a way that illustrates the principles used in acceptor quantum computing. The greatest advantage of acceptors is their high degree of tunability using a gate electric field, which is due to the interplay of the dipole and quadrupole degrees of freedom inherent in the hole spin-3/2. Acceptors in Si combine a series of additional desirable properties. The acceptor confinement potential is built into the device, the sharp confinement helps to reduce decoherence, and its shape is well understood and reproducible. Moreover, devices based on single Si:B acceptors have recently been demonstrated\cite{vanderHeijden:2014fp}, and we expect that fabrication of future Si:B devices will benefit from advances in ion implantation and STM lithography technologies used for fabricating Si:P devices\cite{Morello_1glShot_Nature10,Fuechsle_NNano10}.

Near an interface both heavy (HH) and light hole (LH) spin qubits can be implemented allowing a degree of flexibility. In both architectures all-electrical quantum computing is enabled by the subtle interplay between bulk terms with $T_d$ symmetry and interface-induced Rashba-like spin-orbit terms, which are tunable by a top gate. As compared to Si electrons, holes do not possess a valley degree of freedom, which could interfere with qubit operation and entanglement. The key physics of interface-bound acceptor qubits is conveniently captured by an effective $4 \times 4$ Hamiltonian, which enables an analytical description, and is our central result: 
\begin{equation}\label{GrandH}
H_{eff} = 
\begin{pmatrix}
\varepsilon_{H+} & 0&\varepsilon_\parallel & -i\sqrt{3}pF_z \cr
 0& \varepsilon_{H-} &i\sqrt{3}pF_z & \tilde{\varepsilon}_\parallel \cr
 \varepsilon_\parallel^*&-i\sqrt{3}pF_z & \varepsilon_{L+} &g\mu_B B_- \cr
 i\sqrt{3}pF_z & \tilde{\varepsilon}_\parallel^* & g\mu_B B_+& \varepsilon_{L-}
\end{pmatrix}.
\end{equation}
The basis used here is $\{ 3/2, -3/2, 1/2, -1/2 \}$ in the notation detailed in Sec.~\ref{sec:bulk}. The energies $\varepsilon_{H\pm} = \varepsilon_H \pm \frac{3}{2} \, \varepsilon_Z$, and $\varepsilon_{L\pm} = \varepsilon_L \pm \frac{1}{2} \, \varepsilon_Z$, with $\varepsilon_{H, L}$ the $HH$, $LH$ energies respectively in the absence of magnetic fields, and the Zeeman energy $\varepsilon_Z = g\mu_BB_z$. The terms $\varepsilon_\parallel=-i\sqrt{3}pE_++\frac{\sqrt{3}}{2}g\mu_B B_-$, $\tilde{\varepsilon}_\parallel = -i\sqrt{3}pE_-+\frac{\sqrt{3}}{2}g\mu_B B_+ $ where $E_\pm=E_x\pm iE_y$, $B_\pm=B_x\pm iB_y$ are the in-plane electric and magnetic fields, while $F_z$ is the out-of-plane gate electric field. We provide an analytic model with an analytic solution, with parameters obtained from exact numerical solution to the 6x6 LK Hamiltonian with cubic terms and split-off holes, and parametrized in terms of a series expansion in the gate-electric field. Based on this Hamiltonian and on numerical calculations, we demonstrate that B:Si hole EDSR is especially strong thanks to the interface-induced spin-orbit terms. The biggest challenge facing electrical quantum computation schemes is decoherence due to fluctuating electric fields \cite{Bulaev_PRL05, Fischer_HH_PRL10, Burkard_Decoh_AdvPhys08, Climente_NJP13}. In B:Si, at certain values of the gate electric field we identify sweet spots in the qubit dephasing rates due to charge noise, which are traced to HH-LH mixing, i.e. the quadrupole moment. The EDSR term peaks close to the point where the dephasing time $T_2^*$ also peaks, implying high gate fidelity at the $T_2^*$ sweet spot. Hence acceptor hole qubits can have better coherence times than certain quantum dot-based electron qubits. Entanglement can be accomplished by coupling to a superconducting resonator, using the spin-orbit induced dipole-dipole interaction, or Coulomb exchange \cite{Golovach:2006hx,Flindt:2006dn,Trif:2007fx,Trif:2008cb,Bulaev:2007hj,Kloeffel:2013iv,Ruskov:2013kq,Hu:2012bt,Taylor:2013cq,Blais:2004kn,Wallraff:2004dy,Petersson:2012cv,Xiang:2013hm,Viennot:2015ir,JochymOConnor:2015wp,Salfi:2014kaa,Koiller:2001gw,Samkharadze:2015tw,Graaf:2012fl,Petersson:2010ih}. Our proposed architectures are feasible by combining mature silicon nanofabrication technology with emerging schemes for single-dopant electronics \cite{koenraad2011, Zwanenburg_SiQmEl_RMP13, vanderHeijden:2014fp, Mol:2015im}. 

This paper is part of a series of joint experimental and theoretical studies of acceptor spin qubits. Our experimental work Ref.~\onlinecite{Salfi_2015arxiv} demonstrated the use of acceptor dopants in silicon for quantum simulation, with results backed up by theoretical modeling. Ref.~\onlinecite{Salfi_LH_2015} proposed a spin quantum computation scheme based on the light hole states of an acceptor near an interface (cf. Sec.~\ref{sec:disc}). In this greater context our scope in this paper is twofold: to derive a master Hamiltonian for all interface-bound acceptor qubits and to demonstrate the use of the heavy hole states as a qubit thus completing the greater picture of acceptor spin qubits. 

The outline of this paper is as follows. In Sec.~\ref{sec:bulk} we introduce the total Hamiltonian of an acceptor near an interface. In Sec.~\ref{sec:heavy} we discuss the methodology of implementing a heavy-hole qubit, and a general entanglement platform applicable to all acceptor qubit architectures is introduced in Sec.~\ref{sec:ent}. The heavy-hole and light-hole schemes are compared in Sec.~\ref{sec:disc}, and we end with a summary and conclusions.

\section{Total Hamiltonian}
\label{sec:bulk}

The total Hamiltonian for a boron acceptor located near the Si/SiO$_2$ interface is
\begin{equation}
H = H_{Lut} + H_{T_d} + H_C + H_Z + H_{if} + H_{gt} + H_E
\end{equation}
Here $H_{Lut}$ is the $4 \times 4$ Luttinger Hamiltonian describing holes in bulk semiconductors in the absence of external fields or confinement potential\cite{Luttinger:1955ee}, discussed in detail below. The acceptor Coulomb potential $H_C = e^2/(4\pi\epsilon_0\epsilon_r r)$, with ${\bm r}$ the hole position vector and $\epsilon_r$ the relative permittivity. The Zeeman Hamiltonian $H_Z$ incorporates the interaction with a constant magnetic field, which is needed only to define the qubit. The normal to the interface is taken to be $\parallel \hat{\bm z}$, with Si on the $+z$ side, and the interface potential is represented by $H_{if} = U_0 \Theta(-z)$, with $U_0$ the potential step and $\Theta$ the Heaviside step function. The gate Hamiltonian $H_{gt} = eF_zz$ takes into account the interface electric field $F$, which gives rise to the Rashba terms needed to drive the qubit. The term $H_E$ represents the interaction with the applied time-dependent in-plane electric field ${\bm E}$ for a bulk acceptor. Finally, $H_{Td}$ represents the $T_d$ corrections to the interaction with the electric fields. We explicitly neglect the central cell correction to $H_C$ \cite{Bernholc_PRB77}, which works well for Si:B \cite{Lipari_Acc_78} but which cannot be neglected for other acceptors such as Si:Al, Si:Ga or Si:In.

\subsection{Bulk acceptor}

The Luttinger Hamiltonian is written as
\begin{equation}
\begin{array}{rl}
\displaystyle H_{Lut} = & \displaystyle \bigg(\gamma_1 + \frac{5}{2} \, \gamma_2 \bigg) \, \frac{p^2}{2m_0} - \frac{\gamma_2}{m_0} \, (p_x^2 S_x^2 + p_y^2 S_y^2 + p_z^2 S_z^2) \\ [3ex]
- & \displaystyle \frac{2\gamma_3}{m_0} \, (\{p_x, p_y \}\{S_x, S_y\} + c.p.), 
\end{array}
\end{equation}
where ${\bm p}$ is the momentum operator, $m_0$ the bare electron mass, ${\bm S}$ the effective spin-3/2, and the dimensionless Luttinger parameters for Si are $\gamma_1 = 4.28$, $\gamma_2 = 0.375$, and $\gamma_3 = 1.45$, \cite{Bernholc_PRB77} while $c.p.$ stands for cyclic permutations and $\{A, B\} = AB + BA$. Alternatively, $\mu = (6\gamma_2 + 4\gamma_3)/(5\gamma_1)$ measures the strength of the spherical term, while $\delta = (\gamma_3 - \gamma_2)/\gamma_1 < \mu$ measures the strength of the cubic terms. \cite{Baldereschi_Acc_73, Baldereschi_Acc_74} Since in Si $\mu \approx 0.38$ while $\delta \approx 0.25$, a perturbative treatment of $\delta$ is known to be inaccurate. \cite{Baldereschi_Acc_74, Lipari_Acc_78} In our numerical calculations therefore we have retained the full cubic symmetry of Si and included the spin-1/2 split-off band. Both affect the Si acceptor energy spectrum and are needed for quantitative accuracy, though they do not affect qubit spin dynamics qualitatively. Our analytical exposition therefore relies on the spherical approximation. In the absence of $H_C$, the eigenstates of bulk holes are characterised by the projection of their effective spin-3/2 onto the wave vector, and are referred to as heavy holes (HH, projection $\pm 3/2$) and light holes (HH, projection $\pm 1/2$). For an acceptor in the bulk the ground state is fourfold degenerate, with two states (HH1) predominantly HH and two states (LH1) predominantly LH. \cite{Lipari_Acc_70, Baldereschi_Acc_73, Baldereschi_Acc_74, Lipari_Acc_78, Lipari_SSC80, Song_Acc_EPL13, Masselink_PRB85, Gammon_PRB86, Tezuka_PRB10, Bhatta_PRB72, Gorkov_PRB03, Kopf_PRL92,Smit:2004fx} The states can be labeled by their total angular momentum $\bm{J}=\bm{L}+\bm{S}$, where $\bm{L}$ is a pure orbital operator of the dopant atom. The spin-orbit term couples only states with $\Delta L=0,\pm 2$, defining the wave functions \cite{Baldereschi_Acc_73}
\begin{equation}
\begin{array}{l}
\ket{\Psi_{m_J}} = f_0(r)\ket{L=0,S=\frac{3}{2} ; J=\frac{3}{2}, m_J} \\
\quad\quad\quad+g_0(r)\ket{L=2,S=\frac{3}{2} ; J=\frac{3}{2},m_J} \\ 
\\
\ket{\Phi_{m_J}} = f_2(r)\ket{L=1,S=\frac{3}{2} ; J=\frac{3}{2}, m_J} \\
\quad\quad\quad+g_2(r)\ket{L=3,S=\frac{3}{2} ; J=\frac{3}{2},m_J},
\end{array}
\end{equation}
where $\ket{\Psi_{m_J}}$ are the 4-fold degenerate ground state functions with energy $\varepsilon_0$ and $\ket{\Phi_{m_J}}$ are the first excited states with energy $\varepsilon_1$. The energy eigenfunctions $\ket{\Psi_{m_J}}$ are also eigenstates of $J_z$.

\begin{figure}
\includegraphics[scale=1]{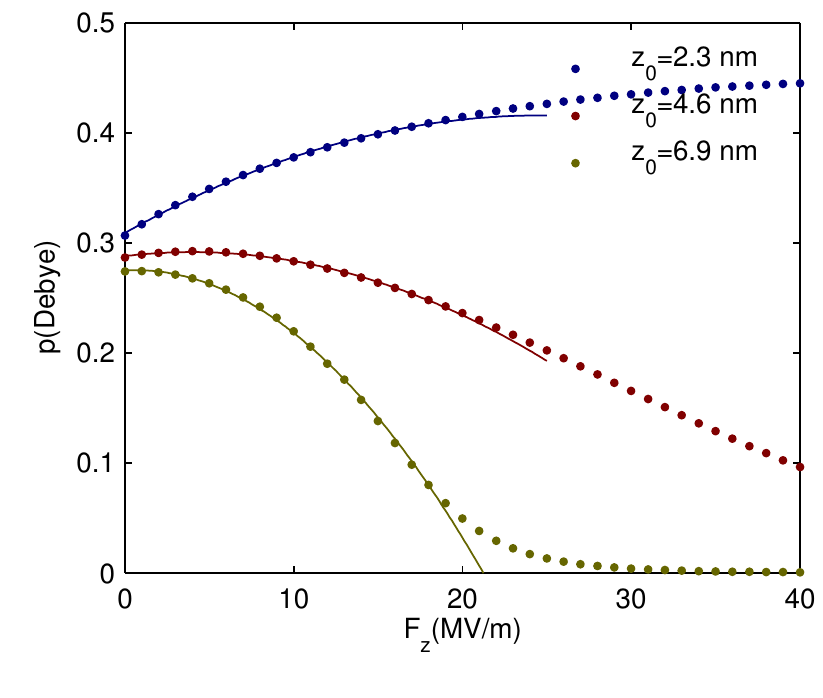}
\caption{Dependence of $p$ on the gate electric field for three acceptor depths $z_0=2.3$ nm, 4.6 nm, and 6.9 nm using a full numerical calculation.}
\label{fP}
\end{figure}

The $T_d$ symmetry of the diamond lattice enables terms linear in the electric field, coupling $HH1$ to $LH1$ \cite{Bir_JPCS63_2}
\begin{equation}
H_{T_d} = p \, E_x \{J_y, J_z\} + c.p.,
\end{equation}
where $p$ is given in Table \ref{tzp} and Fig.~\ref{fP}. \cite{Kopf_PRL92} For spin-3/2 systems interactions with the electric field quadratic in the spin exist which preserve time reversal symmetry. \cite{Bir_JPCS63_2} We shall see below that additional terms $\propto E_xF_z$ exist cf. Refs.~ \onlinecite{Bir_JPCS63_2, Kopf_PRL92}. In general $p$ is a function of the gate field containing even powers of $F_z$ as shown in Fig.~\ref{fP}. It can be written as $p \approx p^{(0)} + p^{(2)}F_z^2$, as in Table~\ref{tzp}, where the coefficient $p^{(1)}$ is negligible. The interface electric field reduces the charge density at the centre of the ion by mixing in higher excited states with $P$ symmetry which have zero charge density at the origin.

\begin{table}
\caption{Coefficients of expansion for $p(F_z)$ defined in the text.}
\begin{ruledtabular}
\begin{tabular}{ l | l | l | l | l }
& Units & & & \\
\hline
$z_0$ & [nm] & 2.3 & 4.6 & 6.9 \\
$p^{(0)}$ & [D]           & $+3.09\times10^{-1}$ & $+2.87\times10^{-1}$ & $+2.75\times10^{-1}$ \\
$p^{(1)}$ & [D(m/MV)]     & $+8.67\times10^{-3}$ & $+1.79\times10^{-3}$ & $+7.70\times10^{-4}$ \\
$p^{(2)}$ & [D(m/MV)$^2$] & $-1.76\times10^{-4}$ & $-2.24\times10^{-4}$ & $-6.46\times10^{-4}$ \\
\end{tabular}
\end{ruledtabular}
\label{tzp}
\end{table}

\subsection{Hamiltonian near an interface}

In the vicinity of the interface, inversion symmetry breaking lifts the degeneracy of the ground state, and provides a natural spin quantisation axis $\parallel \hat{\bm z}$. Symmetry breaking gives rise to a splitting between HH1 and LH1, and to Rashba-like spin-orbit coupling terms.


The interface potential $H_{if}(z)=u_0\Theta(-z)$ is a function of $z=r\cos \theta$ only, and thus does not couple states with different $m_J$. Therefore if we consider only matrix elements within the $4\times4$ bulk ground state manifold, the interface gives only a splitting between HH1 and LH1:
\begin{eqnarray}
u_H=\tbkt{\Psi_{\pm\frac{3}{2}}}{H_{if}(z)}{\Psi_{\pm\frac{3}{2}}}\\
u_L=\tbkt{\Psi_{\pm\frac{1}{2}}}{H_{if}(z)}{\Psi_{\pm\frac{1}{2}}}
\end{eqnarray}
We will use $\varepsilon_{H\pm} = \varepsilon_H \pm \frac{3}{2} \, \varepsilon_Z$, and $\varepsilon_{L\pm} = \varepsilon_L \pm \frac{1}{2} \, \varepsilon_Z$. The energies $\varepsilon_{H, L}$ are the energies in the absence of the Zeeman field: $\varepsilon_H=\varepsilon_0+u_H,\varepsilon_L=\varepsilon_0+u_L$. The splitting between them $\Delta_{HL}=\varepsilon_L - \varepsilon_H = u_L-u_H$ comes from the interface, and has been observed experimentally. \cite{Mol:2015im} The Zeeman energy $\varepsilon_Z = g\mu_BB$, with ${\bm B} \parallel \hat{\bm z}$ for the HH qubit, and ${\bm B} \parallel \hat{\bm x}$ for the LH qubit. The states HH1 are lowest in energy in the absence of strain. \cite{Neubrand:1978je} Henceforth we work in the basis of eigenstates of $H_{Lut} + H_C + H_Z + H_{if}$, and restrict the analytical treatment to the manifold spanned by the four states $\ket{\Psi_{m_J}} \equiv \{3/2,-3/2,1/2,-1/2\}$ corresponding to $HH1$ and $LH1$, which are two sets of Kramers conjugate pairs. 

\begin{figure}
\includegraphics[scale=1]{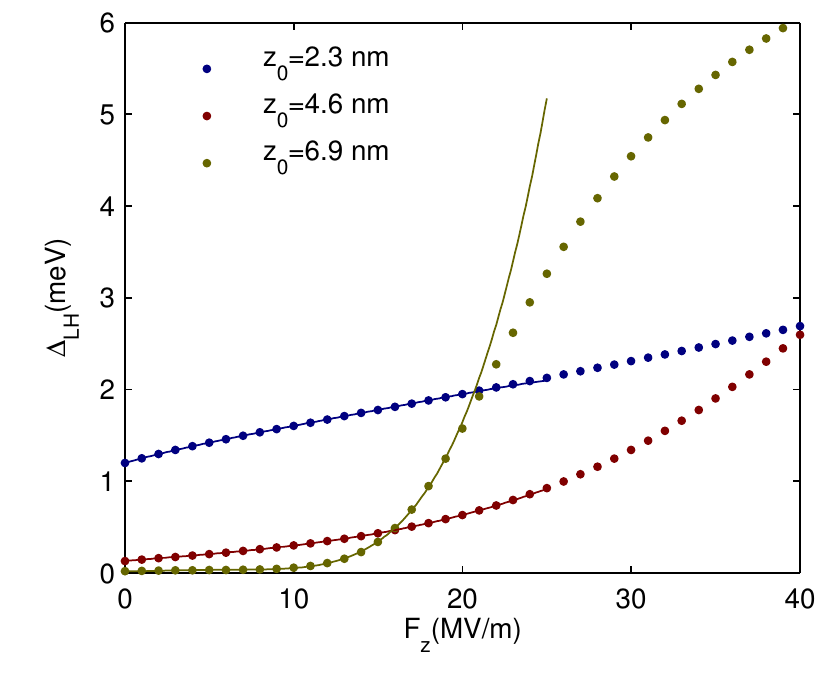}
\caption{Dependence of $\Delta_{\rm LH}$ on gate electric field for three acceptor depths $z_0=2.3$ nm, 4.6 nm, and 6.9 nm.}
\label{fSplitting}
\end{figure}

\begin{figure}
\includegraphics[scale=1]{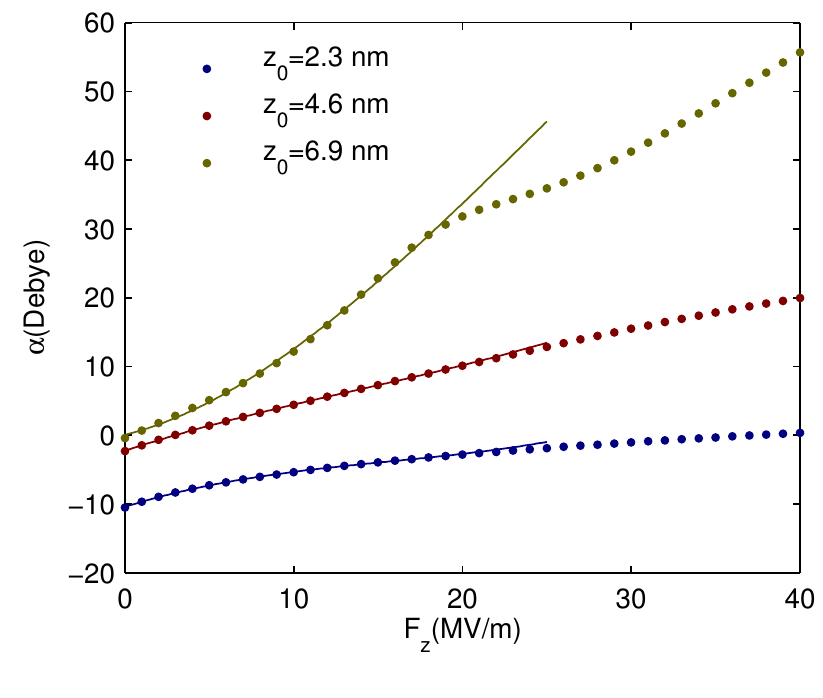}
\caption{Dependence of $\alpha$ on gate electric field for three acceptor depths $z_0=2.3$ nm, 4.6 nm, and 6.9 nm.}
\label{falpha}
\end{figure}

Next we consider the first excited state manifold $\{ \ket{\Phi_{m_J}} \}$ and evaluate Rashba-like spin-orbit terms perturbatively using a Schrieffer-Wolff transformation to project onto the ground state manifold. \cite{Winkler2003} This transformation is described in detail in the supplement. Including $H_{if} + H_{gt}$ and denoting $E_\pm=E_x \pm iE_y$, we find
\begin{equation}
\arraycolsep0.3ex
\begin{array}{rl}
\displaystyle H_R = & \displaystyle 
\alpha \, \begin{pmatrix}
0 & 0& E_- & 0 \cr  
0 & 0 & 0 & - E_+ \cr
E_+ & 0 & 0 & 0 \cr
0 & - E_- & 0 & 0
\end{pmatrix}.
\end{array}
\end{equation}
We can write $\alpha \approx \alpha^{(0)} + \alpha^{(1)} F_z + \alpha^{(2)} F_z^2 + \alpha^{(3)} F_z^3$, where $\alpha^{(0)}$ comes from the interface potential step while the additional terms stem from the interface electric field $F_z$. These terms are given in Table~\ref{talpha}. At low fields we can retain only the linear-in-field term, but at higher values of $F_z$ this approximation does not hold. The full dependence of $\alpha$ on the gate electric field is displayed in Fig.~\ref{falpha}.

\begin{table}
\caption{Coefficients for $\alpha(F_z)$.}
\begin{ruledtabular}
\begin{tabular}{ l | l | l | l | l }
& Units & & & \\
\hline
$z_0$ & [nm] & 2.3 & 4.6 & 6.9 \\
$\alpha^{(0)}$ & [D]           & $-10.4$              & $-2.24$              & $+5.08\times10^{-2}$ \\
$\alpha^{(1)}$ & [D(m/MV)]     & $+7.77\times10^{-1}$ & $+8.04\times10^{-1}$ & $-6.13\times10^{-1}$ \\
$\alpha^{(2)}$ & [D(m/MV)$^2$] & $-3.39\times10^{-2}$ & $-1.76\times10^{-2}$ & $+7.38\times10^{-2}$ \\
$\alpha^{(3)}$ & [D(m/MV)$^3$] & $+7.20\times10^{-4}$ & $+4.22\times10^{-4}$ & $-1.02\times10^{-3}$ \\
\end{tabular}
\end{ruledtabular}
\label{talpha}
\end{table}

We note that Zincblende materials also have Dresselhaus spin-orbit interactions due to bulk inversion asymmetry, \cite{Winkler2003} but these are absent in Si. The Schrieffer-Wolff transformation also yields diagonal terms which give a renormalisation of the splitting between heavy holes and light holes:
\begin{equation}
\begin{array}{c}
\Delta_{HL} = \Delta_{HL}^{(0)}+\Delta_{HL}^{(1)} F_z + \Delta_{HL}^{(2)}F_z^2 + \Delta_{HL}^{(3)}F_z^3 + \Delta_{HL}^{(4)}F_z^4.
\end{array}
\label{RashbaSplitting}
\end{equation}
In a real system higher order terms are always involved, however a parabolic dependence holds as long as $F_z$ is not very large. The coefficients $\Delta_{HL}^{(i)}$ are given in Table~\ref{tdeltaLH}.

\begin{table}
\caption{Coefficients for expansion of $\Delta_{LH}(F_z)$.}
\begin{ruledtabular}
\begin{tabular}{ l | l | l | l | l }
& Units & & & \\
\hline
$z_0$ & [nm] & 2.3 & 4.6 & 6.9 \\
$\Delta_{LH}^{(0)}$ & [eV]           & $+1.20\times10^{-3}$ & $+1.29\times10^{-4}$ & $+1.98\times10^{-5}$ \\
$\Delta_{LH}^{(1)}$ & [eV(m/MV)]     & $+5.10\times10^{-5}$ & $+1.47\times10^{-5}$ & $-4.35\times10^{-6}$ \\
$\Delta_{LH}^{(2)}$ & [eV(m/MV)$^2$] & $-1.77\times10^{-6}$ & $-9.83\times10^{-8}$ & $+3.55\times10^{-6}$ \\
$\Delta_{LH}^{(3)}$ & [eV(m/MV)$^3$] & $+8.65\times10^{-8}$ & $+3.11\times10^{-8}$ & $-5.97\times10^{-7}$ \\
$\Delta_{LH}^{(4)}$ & [eV(m/MV)$^4$] & $-1.58\times10^{-9}$ & $-1.31\times10^{-11}$ & $-3.16\times10^{-8}$ \\
\end{tabular}
\end{ruledtabular}
\label{tdeltaLH}
\end{table}

\subsection{Our approach}

In this work we use a hybrid approach combining analytics and numerics, which gives both physical insight and numerical accuracy. The analytical approach is based on the simplified $4\times 4$ Luttinger Hamiltonian in the spherical approximation introduced above, in which the cubic terms and the spin split-off band are neglected. We include the $T_d$ interaction terms with the electric fields as well as the interface-induced level splittings and spin-orbit terms, and adding up all the terms introduced above we obtain the Hamiltonian of Eq.\ (\ref{GrandH}). The numerical approach is a non-perturbative, exact diagonalisation $6 \times 6$ Luttinger-Kohn calculation explicitly including the cubic terms, the ion and interface well potentials and all electric field terms. We account for hybridisation of levels outside the $4 \times 4$ extended qubit subspace, leading to renormalised parameters for e.g. the Zeeman interactions. Since the analytics cannot yield the coefficients e.g. $\alpha^{(0)}$ and $\alpha^{(1)}$ for the interface spin-orbit coupling, these coefficients are obtained by fitting the analytical results to the numerics, thus we keep all the terms in the analytics required to reproduce the numerical results.

\section{Heavy Hole Qubit}
\label{sec:heavy}

Taking Eq.\ (\ref{GrandH}) as our starting point, we now describe in detail the operation of a HH qubit encoded in the effective $HH1$ Kramers doublet $\ket{\Psi_{\pm\frac{3}{2}}}$. A constant uniform Zeeman field $\parallel \hat{\bm z}$ separates the qubit states, such that a hole can be loaded into the $m_J = - \frac{3}{2}$ state in the HH1 manifold. A Zeeman field of $\approx 0.5$ T produces an energy splitting of $\approx 50\mu$eV $\gg k_BT$ in a dilution refrigerator, hence initialisation is straightforwardly accomplished. The in-plane magnetic field in Eq.\ (\ref{GrandH}) is set to zero for the remainder of this discussion.

\subsection{Effective HH Qubit Hamiltonian and states}

We assume $E_x, E_y \ll F_z$ and initially we take $E_x=E_y=0$. Diagonalizing Eq.\ (\ref{GrandH}), assuming $\Delta_{HL}>2\varepsilon_Z$, we obtain the eigenvalues
\begin{equation}
\begin{array}{rl}
\lambda_{H+}&=\dfrac{1}{2}[(\varepsilon_{H+}+\varepsilon_{L-})-\sqrt{(\varepsilon_{H+}-\varepsilon_{L-})^2+12p^2F^2}]\\
\lambda_{H-}&=\dfrac{1}{2}[(\varepsilon_{H-}+\varepsilon_{L+})-\sqrt{(\varepsilon_{H-}-\varepsilon_{L+})^2+12p^2F^2}]\\
\lambda_{L+}&=\dfrac{1}{2}[(\varepsilon_{H-}+\varepsilon_{L+})+\sqrt{(\varepsilon_{H+}-\varepsilon_{L-})^2+12p^2F^2}]\\
\lambda_{L-}&=\dfrac{1}{2}[(\varepsilon_{H+}+\varepsilon_{L-})+\sqrt{(\varepsilon_{H-}-\varepsilon_{L+})^2+12p^2F^2}]
\end{array}
\end{equation}
with corresponding eigenvectors
\begin{equation}
\arraycolsep0.3ex
\begin{array}{rl}
\displaystyle \ket{v_{H+}} = & \displaystyle \frac{1}{N_1} \begin{pmatrix} -\frac{i\sqrt{3}pF}{(\lambda_{H+} - \varepsilon_{H+})} \cr 0 \cr 0 \cr 1
\end{pmatrix} = \frac{1}{N_1} \begin{pmatrix} -ia \cr 0 \cr 0 \cr 1
\end{pmatrix}  \\ [3ex]
\displaystyle \ket{v_{L-}} = & \displaystyle \frac{1}{N_1} \begin{pmatrix} 1 \cr 0 \cr 0 \cr -\frac{i\sqrt{3}pF}{(\lambda_{H+} - \varepsilon_{H+})} 
\end{pmatrix} = \frac{1}{N_1} \begin{pmatrix}  1 \cr 0 \cr 0 \cr -ib 
\end{pmatrix} \\ [3ex]
\displaystyle \ket{v_{H-}} = & \displaystyle \frac{1}{N_2} \begin{pmatrix} 0 \cr \frac{i\sqrt{3}pF}{(\lambda_{H-} - \varepsilon_{H-})} \cr 1 \cr 0
\end{pmatrix} = \frac{1}{N_2} \begin{pmatrix} 0 \cr ib \cr 1 \cr 0
\end{pmatrix} \\ [3ex]
\displaystyle \ket{v_{L+}} = & \displaystyle \frac{1}{N_2} \begin{pmatrix} 0 \cr 1  \cr \frac{i\sqrt{3}pF}{(\lambda_{H-} - \varepsilon_{H-})} \cr 0
\end{pmatrix} = \frac{1}{N_2} \begin{pmatrix} 0 \cr 1 \cr ib \cr 0
\end{pmatrix}.
\end{array}
\label{eigenvector}
\end{equation}
where $N_1, N_2$ are normalisation factors, given by:
\begin{equation}
N_{1,2} = \sqrt{1 + \frac{3p^2F^2}{(\lambda_{H\pm} - \varepsilon_{H\pm})^2}}.
\end{equation}
The corresponding eigenvector matrix $V$ is given in the supplement. The originally 4-fold degenerate subspace was first split into HH and LH subspaces as a result of the interface potential, each being doubly degenerate. This remaining degeneracy is completely removed in response to the applied Zeeman field. In the presence of the Zeeman field, the energy of the state $\ket{u_{H+}}$ is lifted while the energy of $\ket{u_{L-}}$ is suppressed. We notice that the two level cross each other at $\varepsilon_Z=\frac{1}{2}\Delta_{HL}$, thus $\frac{1}{2}\Delta_{HL}$ is the critical Zeeman field strength below which the two lowest energy levels are the heavy holes $\ket{u_{H\pm}}$ and above which are $\ket{u_{H-}}$ and $\ket{u_{L-}}$. Without the interface spin-orbit terms
\begin{equation}
\arraycolsep0.3ex
\begin{array}{rl}
{\displaystyle H_{qbt}^{(0)} =} & {\displaystyle \begin{pmatrix}
\varepsilon_{H+} & 0 & -i\sqrt{3} p E_+ & -i\sqrt{3} pF_z \\
0 & \varepsilon_{H-} & i\sqrt{3} pF_z  & -i\sqrt{3} p E_- \\
i\sqrt{3}pE_- & -i\sqrt{3}pF_z & \varepsilon_{L+} & 0 \\
i\sqrt{3} pF_z & i\sqrt{3}pE_+ & 0 & \varepsilon_{L-}
\end{pmatrix}}.
\end{array}
\label{eq:eff}
\end{equation}

\begin{figure}
\includegraphics[scale=1]{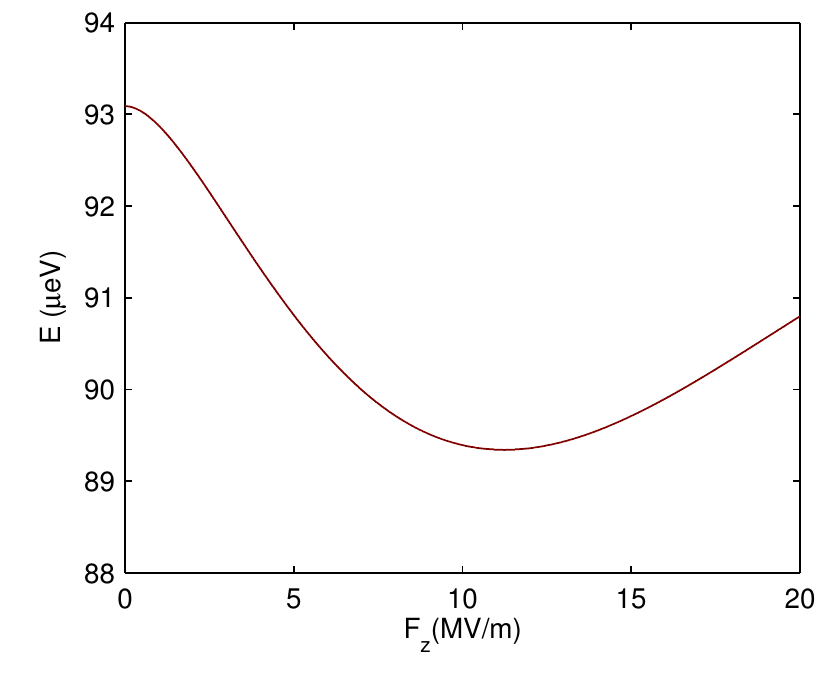}
\caption{Dependence of qubit splitting on gate electric field for $z_0=4.6$ nm.}
\label{flarmor}
\end{figure}

The dependence of the qubit Larmor frequency on the gate electric field is shown in Fig.~\ref{flarmor}.

\subsection{EDSR}
\label{sec:EDSR}

\begin{figure}
\includegraphics[scale=1]{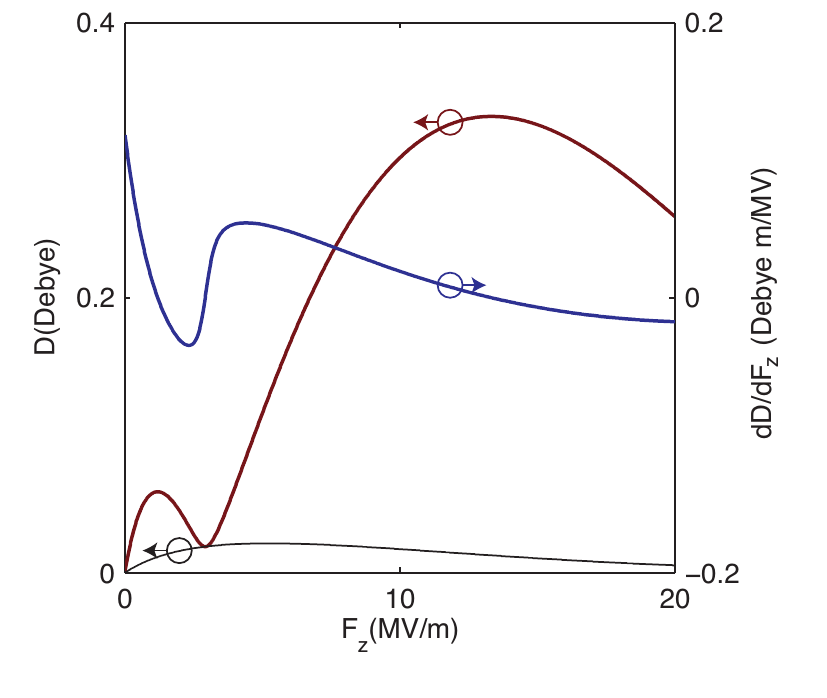}
\caption{Dependence of EDSR coupling $D$ on gate electric field for $z_0=4.6$ nm, including the interface term (red) and without the interface term (black). The sensitivity of gate timing to electric field noise is proportional to the first derivative of $D$ with respect to electric field, $dD/dF_z$ (blue).}
\label{fEDSR}
\end{figure}

\textit{Rotated Rashba terms}: The off-diagonal terms of Rashba Hamiltonian couple between heavy holes and light holes. In the basis $\{\ket{v_{H+}},\ket{v_{H-}},\ket{v_{L+}},\ket{v_{L-}}\}$ the off-diagonal matrix is transformed to 
\begin{equation}
\arraycolsep0.3ex
\begin{array}{rl}
\displaystyle \tilde{H}_R = & \displaystyle \alpha \begin{pmatrix}
0 & -i\xi_1E_-  & \xi_2 E_-  & 0 \cr
i\xi_1 E_+  & 0 & 0 & -\xi_2 E_+ \cr
\xi_2 E_+  & 0 & 0 & -i\xi_1 E_+  \cr
0 & -\xi_2 E_-  & i\xi_1 E_-  & 0 
\end{pmatrix}
\end{array}
\label{pRashba}
\end{equation}
where $\xi_1 = (a - b)/(N_1N_2)$, $\xi_2 = (1 + ab)/(N_1N_2)$. This Hamiltonian describes the response of the HH qubit to the in-plane electric field through the Rashba interaction. It has an overall dependence on $F_z$ which is the origin of all the novel physics in the vicinity of this interface. In this same basis
\begin{equation}
\arraycolsep0.3ex
\begin{array}{rl}
\displaystyle \tilde{H}_E = & \displaystyle \sqrt{3}p \begin{pmatrix}
0 & \xi_1E_+ & i\xi_2 E_+ & 0 \\
\xi_1E_- & 0 & 0 & i\xi_2E_-\\
-i\xi_2E_- & 0 & 0 & -\xi_1E_-\\
0 & -i\xi_2E_+ & -\xi_1E_+ & 0
\end{pmatrix}.
\end{array}
\label{peff}
\end{equation}
We further project $\tilde{H}_E$ into the HH subspace using a Schrieffer-Wolff transformation, whence the original dipole moment $p \rightarrow p_{eff}=\sqrt{3}\xi_1 p$. We apply the same transformation to $\tilde{H}_R$, finding that the Rashba-induced dipole moment $p_R \approx -i[\alpha^{(0)}+\alpha^{(1)}F_z]\xi_1$, thus the total EDSR dipole moment is $D = |p_{eff}+p_R|$, with the interface-induced spin-orbit terms making a significant contribution. The dependence of $D$ on the gate electric field is shown in Fig.~\ref{fEDSR}. The blip in this figure indicates the point at which the interface-induced spin orbit terms become dominant over the bulk $T_d$-symmetry $p$-terms.

\subsection{Relaxation}

\begin{figure}
\includegraphics[width=\columnwidth]{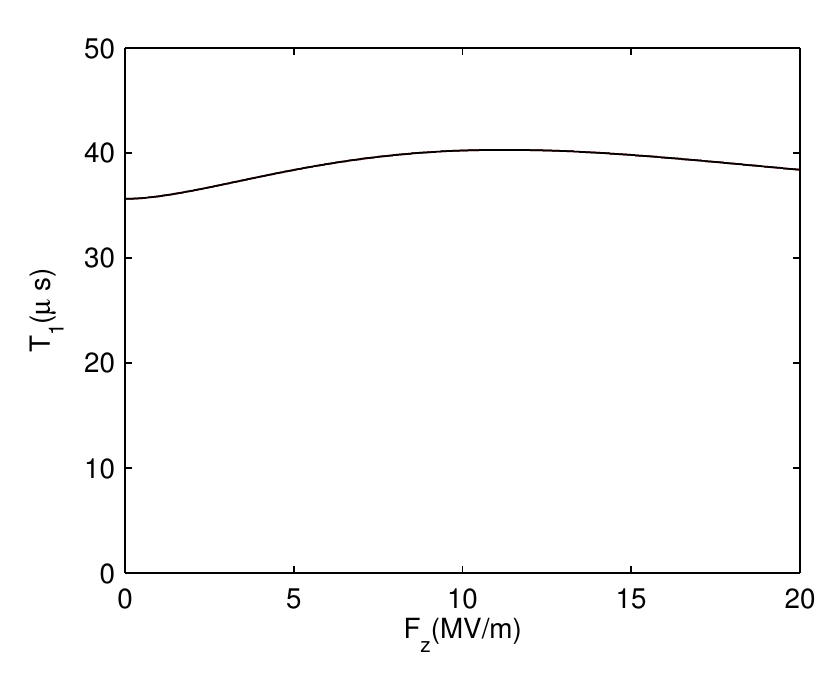}
\caption{Dependence of $T_1$ on gate electric field for $z_0=4.6$ nm and $B=0.5$ T.}
\label{HeavyRelaxation}
\end{figure}

Qubit relaxation is expected to be due predominantly to acoustic phonons. For the heavy hole qubit the spin relaxation time is given by the standard expression
\begin{equation}
\frac{1}{T_1}=\frac{(\hbar\omega)^3}{20\hbar^4\pi\rho}\Big(\frac{\varepsilon_Z}{\tilde{\Delta}_{HL}}\Big)^2\Big[6d'^2\Big(\frac{2}{3v_l^5}+\frac{1}{v_t^5}\Big)\Big],
\end{equation}
where $\varepsilon_Z=g\mu_BB$, $\hbar\omega$ is the qubit frequency, $\tilde{\Delta}_{HL}$ is the splitting between heavy hole subspace and light hole subspace modified both by the interface and gate electric field, which is given in the supplement, where the details of the derivation are also provided. This value of $T_1^{-1}$ is approximately $3(\varepsilon_Z/\tilde{\Delta}_{HL})^2$ smaller than the value for the $\left|-1/2\right\rangle$ to $\left|-3/2\right\rangle$ transition in Si:B.  The factor of 3 comes because the Zeeman energy is tripled, and the factor $(\varepsilon_Z/\tilde{\Delta}_{HL})^2$ comes because the transition here is for a spin qubit. We use $v_l=8.99\times10^{3}$ m/s and $v_t=1.7v_l$ for the longitudinal and transverse sound velocities in silicon, respectively, $\rho=2330$ kg/m$^3$ for the mass density, and $b'=-1.4$ eV and $d'=-3.7$ eV for the Bir-Pikus deformation potentials.  We obtain $T_1=40$ $\mu$s for $B=0.5$ T, and $T_1=1.3$ ms for $B=0.25$ T. The dependence of $T_1$ on the gate electric field is shown in Fig.~\ref{HeavyRelaxation}.

\subsection{Dephasing}

\begin{figure}
\includegraphics[width=\columnwidth]{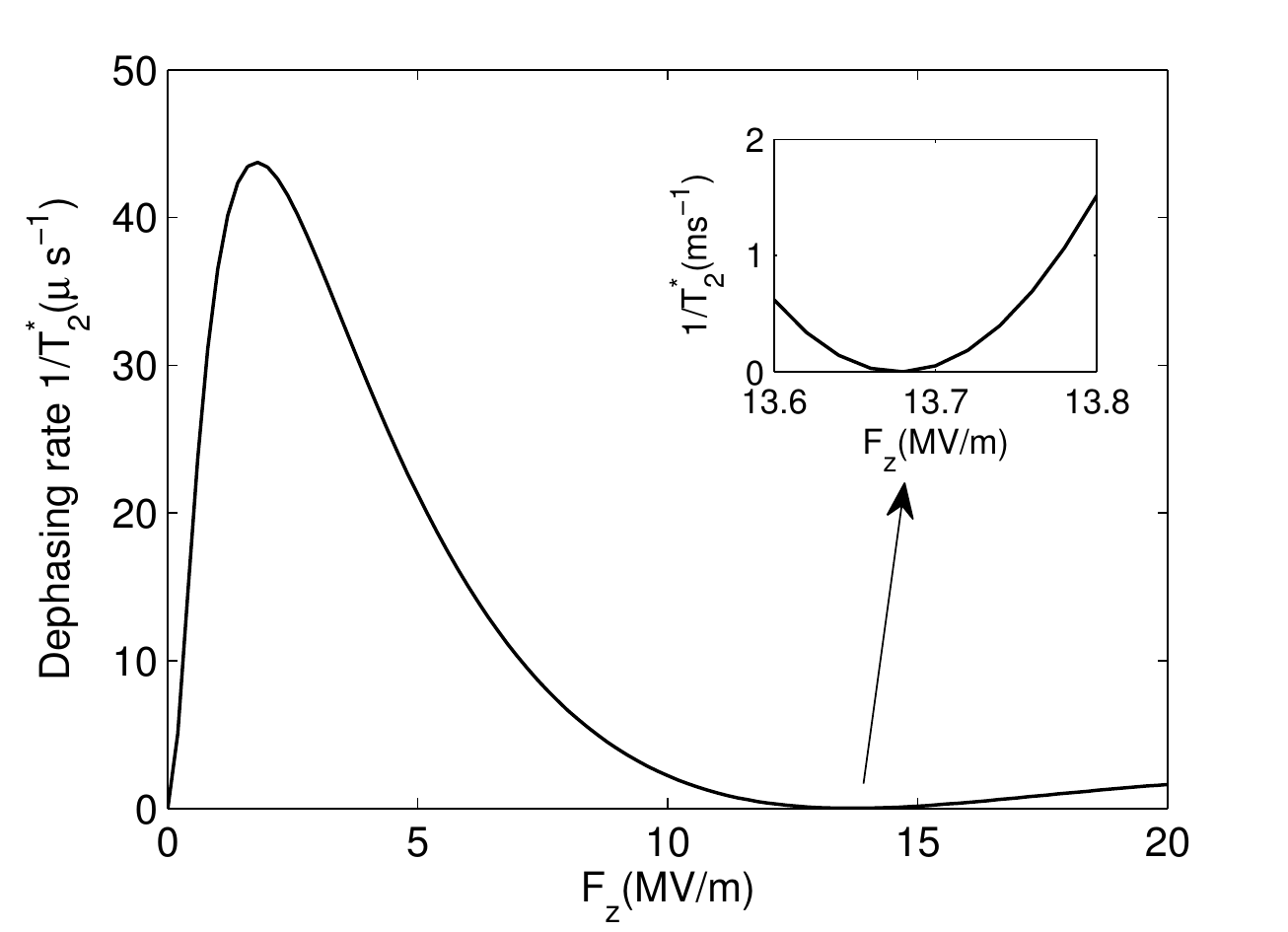}
\caption{Dependence of $T_2^*$ on the gate electric field $F_z$ for $z_0=4.6$ nm using an analytical approximation. The sweet spot occurs at $F_z=13.68$ MV/m in this approximation, which is marginally different from the numerical result.}
\label{HeavyDephasing}
\end{figure}

For a Zeeman energy $\varepsilon_Z<\frac{1}{2}\Delta_{HL}$, the two lowest energy states are $\ket{v_{H-}}$ and $\ket{v_{H+}}$. To analyse the promise of constructing a highly coherent qubit out of these two heavy hole states, we evaluate the free induction decay dephasing time $T_2^*$. \cite{Bermeister_APL14, Sousa_PRB03, Hu_PRL06, Burkard_Decoh_AdvPhys08, Culcer_APL09, Sousa_BookChapter_09, RamonHu_DQD_Decoh_PRB10, Culcer_APL13} Common sources of dephasing in solid state spin qubits include the hyperfine interaction, fluctuations in the Land\'e $g$-factor and through the joint effect of hyperfine or spin-orbit coupling and fluctuating
electric fields due to phonon effects or charge noise. \cite{Merkulov_PRB02, Khaetskii_PRB03, Deng_PRB06, Coish_Nuclear_PSSB09, Ivchenko_QW_g_fluct_SSC97, Erlingsson_2002_hyperfine, Golovach_PhnDecay_PRL04, Bulaev_PRL05, Schoen_Dephasing_QD_PRL06, Prada_PRB08,Hu_2spin_e-ph_PRB11, Climente_NJP13, Fleetwood, Jung_APL04, PhysRevB.89.195302} In silicon the hyperfine effect can be removed by isotopic purification while phonon effects decrease in importance at low magnetic fields, therefore the susceptibility to charge noise is expected to be the main source of dephasing. \cite{Yacoby_Nuclear_Bath_PRL10, Witzel_AHF_PRB07, Petersson_PRL10, PhysRevLett.110.136802, Dial_ST_PRL13, Paladino_1/f_RMP14} We introduce explicitly the vertical electric field $\mathcal{F}_z$ due to the defect, which modifies the total vertical electric field as $F_z \rightarrow F_z + \mathcal{F}_z$, and the HH-LH splitting as
\begin{equation}
\Delta_{HL} \rightarrow \Delta_{HL}^{(0)}+\Delta_{HL}^{(1)}(F_z + \mathcal{F}_z)+\Delta_{HL}^{(2)}(F_z + \mathcal{F}_z)^2.
\end{equation} 
We will retain only terms up to first order in $\mathcal{F}_z$, since in realistic samples active defects will be some distance $d$ away from the acceptor. We will assume $d = 40$ nm. 
The dephasing rate is plotted in Fig.~\ref{HeavyDephasing}, in which a sweet spot is notable. We find $T_2^* > 1$ ms for $F_z$ within $\pm 0.1$ MV/m of the sweet spot, which is more than $10^3$ times the value of the obtained entanglement gate times. To obtain these figures, firstly a realistic defect separation $d = 40$ nm is considered, since the immediate vicinity of the acceptor is depleted by the gates. The value of the defect $\mathcal{F}_z$ is 3380 V/m using $\epsilon = 11.8$, and a defect switching time $\tau = 1000 \tau_1$ is considered as a worst case estimate. Here $\tau_1$ is taken as the EDSR gate time for a $\pi$-pulse represented by $h/2DE_{ac}$, with $E_{ac}$ the ac in-plane electric field, estimated as 500 V/cm, and the dipole moment $D$ estimated at the sweet spot from Fig.~\ref{fEDSR}. This choice is justified by the fact that the effect of slower fluctuators can be strongly suppressed by dynamical decoupling pulses, which in our architecture are mediated by EDSR. 

Importantly, we find that the peak in $T_2^*$ occurs at a point where the EDSR strength is near the maximum, which implies that at the $T_2^*$ sweet spot the EDSR term exhibits very little sensitivity to noise. Since very little jitter in the gating is expected at this point well-defined gate times can be obtained as well as high fidelity. The same may be true of the circuit quantum electrodynamics (cQED) and dipole-dipole entanglement schemes discussed below, which rely on the EDSR gate to couple the qubit spin to an electric field. 

\section{Entanglement} 
\label{sec:ent}

The transverse coupling of both the HH and LH qubits to electric fields allows for entanglement of spins over long distances. Two schemes of immediate interest for spin-spin coupling include electric dipole-dipole interactions, or cQED using superconducting microwave cavities. 

Long-range schemes for entanglement have important advantages over short-range schemes such as Heisenberg exchange, because they reduce qubit density, reducing the density of qubit control lines, and make entanglement operations far less susceptible to errors in positioning of the dopants. Acceptor-bound holes are ideal for long-distance coupling of spins mediated by electric fields and photons, since their large spin-orbit coupling allows for strong couplings to electric fields resulting in fast operations over long distances. Moreover, despite the strong coupling to electric fields, the spins are protected at the Hamiltonian level against electric field noise, when operating at the sweet spot during entanglement operations.  

\subsection{Dipole-dipole interactions}

Given that Coulomb interactions couple into the spin-subspace due to spin-orbit coupling, two qubit gates can be achieved by electric dipole-dipole interactions \cite{Flindt:2006dn,Trif:2007fx}. We consider the mutual Coulomb interaction for qubits with $R_{12}\gtrsim 20$ nm separation, where the exchange interaction is sufficiently small to be neglected.  

The Hamiltonian in the two-spin subspace is $H=H_1+H_2+H_{\rm int}$ where $H_i$ is the extended $4\times4$ Hamiltonian for isolated qubit $i=1$ and $2$, and $H_{\rm int}$ is their mutual Coulomb interaction.  Working in a $16\times 16$ direct product subspace $\ket{pq}=\ket{v_{p}}\otimes\ket{v_{q}}$ where $p\in\{H\pm,L\pm\}$ and $q\in\{H\pm,L\pm\}$, the dipole-dipole interaction in the multi-pole expansion is
\begin{align}
\bra{pq}V^{12}\ket{p'q'}=\frac{e^2}{4\pi\epsilon R_{12}^5}\Big[R_{12}^2\bkt{\delta\mathbf{r}_1}_{pp'}\cdot\bkt{\delta\mathbf{r}_2}_{qq'}&\nonumber\\-3(\bkt{\delta\mathbf{r}_1}_{pp'}\cdot\mathbf{R}_{12})(\bkt{\delta\mathbf{r}_2}_{qq'}\cdot\mathbf{R}_{12})&\Big]
\end{align}
where we have introduced
\begin{equation}
\bkt{\delta\mathbf{r}_i}_{pp'}=\int dr^3_i (\mathbf{r}_i-\mathbf{R}_i) v_p^\dagger(\mathbf{r}_i)v_{p'}(\mathbf{r}_i).
\end{equation}
That is, $\bkt{\delta\mathbf{r}_i}_{pp'}$ is related by scale factors to the sum of quadrupolar interactions of individual qubits with electric fields, $\tilde{H}_{Rnn'}+\tilde{H}_{Enn'}$, given by Equation (\ref{pRashba}) and Equation (\ref{peff}).  Here, $\mathbf{R}_i$ is the position of qubit ion $i$, and $\mathbf{R}_{12}=\mathbf{R}_1-\mathbf{R}_2$.  

The combined effect of Zeeman and Coulomb perturbations can be projects into the coupled qubit subspace $\ket{H-H-}, \ket{H-H+}, \ket{H+H-}, \ket{H+H+}$ as a spin-dependent interaction, using a Schrieffer-Wolff transformation.  This interaction, which is physically interpreted as Ising-type interaction, has a magnitude $J_{dd} = [\mathbf{v}_1 \cdot \mathbf{v}_2 R_{12}^2 - 3(\mathbf{v}_1\cdot\mathbf{R}_{12})(\mathbf{v}_2 \cdot \mathbf{R}_{12})]/(4\pi \epsilon R_{12}^5)$, where $\mathbf{v}_i$ is the spin-dependent charge dipole of qubit $i=1,2$. After some simplification we obtain $J_{dd}=D^2/4\pi \varepsilon_0 \varepsilon_r R_{12}^3$.  An electric dipole-dipole mediated two-qubit $\sqrt{\rm SWAP}$ time of $\tau_{dd}=h/4J_{dd} = 1$ $\mu$s is obtained. 

\subsection{Circuit QED}

For entanglement using cQED it is customary to start from the well-known Jaynes-Cummings Hamiltonian
\begin{equation}
H_{JC} = \hbar \omega_r (a^\dag a + 1/2) + \frac{\hbar\omega}{2} \, \sigma_z + \hbar g_c (a^\dag \sigma_- + \sigma_+ a)
\end{equation}
where $a$, $a^\dag$ represent photon creation and annihilation operators respectively for a resonator with frequency $\omega_r$, $\omega$ is the qubit frequency, and $g_c = eDE_0/\hbar$ the vacuum Rabi coupling constant, given by the product of the qubit dipole matrix element and the vacuum electric field $E_0$ of the resonator mode. 

By means of a unitary transformation, described in Ref.~\onlinecite{Blais:2004kn}, the following effective two-qubit (1,2) interaction Hamiltonian can be derived
\begin{equation}
\begin{array}{rl}
\displaystyle H = & \displaystyle [\hbar\omega_r + \frac{\hbar g_c^2}{\Delta} \, (\sigma_z^{(1)} + \sigma_z^{(2)})] \, a^\dag a \\ [1ex]
+ & \displaystyle \bigg(\frac{\hbar\omega}{2} + \frac{\hbar g_c^2}{\Delta}\bigg) \, (\sigma_z^{(1)} + \sigma_z^{(2)}) \\ [1ex]
+ & \displaystyle \frac{\hbar g_c^2}{\Delta} \, (\sigma_+^{(1)} \sigma_-^{(2)} + \sigma_-^{(1)} \sigma_+^{(2)}).
\end{array}
\end{equation}
Here the detuning $\Delta = \omega - \omega_r$ is the difference between the Larmor energy of the qubit $\hbar\omega$ and the photon energy $\hbar\omega_r$ in the resonator. This indicates that an effective spin-spin interaction of magnitude $J_{c}=2\hbar g_c^2/\Delta$ can be achieved, whose magnitude is estimated as follows. 

Assuming a coplanar waveguide resonator operating at $B=0.5$ T narrowed to a gate of 60 nm in the vicinity of the dopant, a vacuum electric field of $E_0=50$ V/m should be achievable, and the vacuum Rabi coupling $g_c=eDE_0/\hbar$ can be found from Fig.~\ref{fEDSR} by taking the value of $D$ at the sweet spot. Operating at a detuning of $\Delta=4g_c$, we obtain a two-qubit $\sqrt{\rm SWAP}$ time of $\tau_{c}=h/4J_{c,xx}=6.6$ $\mu$s. Recently, quality factors exceeding $10^5$ have been achieved for fields up to $0.5$ T in an in-plane field geometry, which corresponds to a photon decay rate of approximately $\kappa\sim(40$ $\mu\textrm{s})^{-1}$.

We note that the insensitivity of the qubit Larmor frequency to charge noise also protects the qubits against decoherence during two-qubit gates.  Moreover, the strength of the dipole-dipole and cQED mediated interactions is proportional to $D^2$, such that the jitter in gate times goes as $dD^2/dF_z$, see Fig.~\ref{fEDSR}.  Here we see that $dD^2/dF_z$ goes to zero near the sweet spot, also helping to minimize gate errors that might be introduced by charge noise during two-qubit operations.

\section{Discussion}
\label{sec:disc}

The quadrupolar spin-orbit interactions described herein are general to acceptor-bound holes experiencing additional confinement from an interface and a gate electric field. For example, they are the essential ingredients for the proposed electrically controllable light-hole qubit in Ref.~\onlinecite{Salfi_2015arxiv}. They include the (i) $T_d$ symmetry electric-field interactions, (ii) interface-induced splitting, and (iii) interface induced spin-orbit coupling. 

For both the HH qubit described herein, and the LH qubit in Ref.~\onlinecite{Salfi_2015arxiv}, the gate-electric field is used to tune the interface-induced splitting and the interface-induced spin-orbit coupling between the light and heavy holes. Moreover, in both cases, the $T_d$ symmetry spin-orbit interaction with gate electric fields mixes the light and heavy holes. This mixing is responsible for two effects: first, together with the applied magnetic field, it transforms the spin-orbit coupling between the light and heavy holes into a transverse EDSR term in the qubit subspace, and second, it introduces a dispersion into the ground state that gives rise to the sweet spot, which makes the qubit insensitive to electric field noise. We emphasise that both qubits can be described starting from the generic Hamiltonian introduced in Eq.~(\ref{GrandH}), which is thus regarded as a master Hamiltonian of acceptor spin qubits. The key ingredients enabling full electrical control of acceptor qubits are the interface-induced spin-orbit terms, which may be regarded as the spin-3/2 equivalent of the Rashba terms known in semiconductor nanostructures. These fall into two categories: a contribution due to the potential energy step at the interface, which has not been calculated before, and a contribution due to the interface electric field, which can be identified with the $d$-terms in Ref.~\onlinecite{Bir_JPCS63_2}. 

In spite of the overall similarities, the heavy-hole qubit that has been the focus of this paper is described by a qualitatively different effective Hamiltonian than the one in Ref.~\onlinecite{Salfi_2015arxiv}. This is because it relies on the spin-diagonal Zeeman interaction with an out-of-plane magnetic field, since HH have vanishing Zeeman interactions for in-plane magnetic fields. On the other hand, the light-hole qubit of Ref.~\onlinecite{Salfi_2015arxiv} has spin off-diagonal Zeeman interactions because the applied magnetic field is perpendicular to the interface-induced spin quantization axis. The in-plane $g$-factor can be sizeable thanks to couplings between the heavy holes and the light holes. In part due to these qualitative differences, the physical origin of the sweet spot in $T_2^*$ is different for the two systems. For the LH qubit with an in-plane magnetic field a very generic reason for the sweet spot can be identified, namely the non-perturbative HH-LH mixing near the anti-crossing, where the qubit states comprise a $\sqrt{3}/2$ spectral amplitude of light holes and a $1/2$ spectral amplitude of heavy holes. Moreover, for an out of plane applied magnetic field, the LH qubit architecture would not exhibit a sweet spot in $T_2^*$. The sweet spot for the HH qubit, on the other hand, is not liable to a simple explanation. It is absent if $\Delta_{HL}^{(2)}$ is artificially set to zero, even for $p \ne 0$, which is not a restriction on the realization of the sweet spot for the LH qubit.

The two qubits we have introduced in this work and in Ref.~\onlinecite{Salfi_2015arxiv} can be controlled solely by electrical means. Quite generally in both cases, the location of the qubit with respect to the interface is important in achieving a high degree of electrical tunability. Having the qubit too close to the interface is not ideal because the HH-LH splitting $\Delta_{HL}$ becomes very large, suppressing HH-LH mixing terms. At the same time it is undesirable to have the qubit too far from the interface, since the $eF_zz$ term increases linearly with $z$. The ground state is a hybrid between triangular and spherical eigenstates and the anti crossing between the two becomes very sharp (narrow) at large values of $eF_zz$. \cite{Calderon_PRB08} This latter shortcoming can be compensated using strain. \cite{Salfi_2015arxiv,vanderHeijden:2014fp,Lo:2015kh,Celler:2003ha}

Two notable quantitative differences stand out between the HH and LH qubits. Firstly, as compared to the LH qubit, the effective EDSR dipole moment for the HH qubit is weaker by a factor of approximately 30. This is compensated by the fact that for the exact same physical setup the HH qubit is considerably less sensitive to noise than the LH qubit, exhibiting much longer dephasing times around the sweet spot. Furthermore, the second order term in the noise electric field is expected to be much smaller for the HH qubit because the curvature of the Larmor frequency as a function of gate electric field is considerably smaller. This fact, and the quantitative description of dephasing, will be addressed in detail in a future publication. 

Overall, the LH qubit has more extreme performance characteristics, since it allows for larger drives to be obtained, giving more gate operations per $T_1$. However, it is harder to make, requiring strain.  Moreover, obtaining larger drive strengths requires deeper acceptors, such that the sweet spot has a narrower window of operation where $T_2^*\gg T1$. The HH qubit on the other hand allows fewer gate operations per $T_1$, since around the sweet spot $T_2$ is limited by $T_1 \approx 100$ $\mu$s, allowing 100 qubit operations compared with $10^4$ for the LH qubit. However, it has a much larger window in which the charge noise $T_2^* \gg T_1$, and it does not require strain, simplifying the fabrication. At the same time, for the HH qubit, to a first approximation the Zeeman interaction with an in-plane magnetic field vanishes, therefore a magnetic field perpendicular to the interface is needed to split the qubit states. Such a perpendicular field may affect the $Q$-factor of a superconducting resonator used in entanglement via circuit QED. In that case it will be preferable to use dipolar coupling for entanglement, a scheme for which we have found no obvious restrictions. 

Finally, we note that two-qubit gates, mediated by dipole-dipole interactions or cQED and producing entanglement, can either be turned ``on'' and ``off'', by varying the gate electric field, or alternatively, ``always-on'' interactions could be employed\cite{Zhou:2002jg}.  The ``on'' behaviour is obtained at the finite-field sweet spot, where EDSR-mediated cQED and dipole-dipole interactions are strong.  The ``off'' behaviour is obtained by adiabatically sweeping at the $F_z\sim 0$ (zero-field) sweet spot where EDSR is considerably weaker, see Fig.~\ref{fEDSR}.  Although couplings $\alpha E_{x,y}$ between the light and heavy holes are present at $F_z\sim 0$ due to the interface, the $pF_z$-induced light-heavy hole mixing vanishes, and as such, the $\alpha E_{x,y}$ interactions do not couple into the qubit subspace as EDSR interactions.  Alternatively, several schemes exist which can provide a universal set of operations, without the ability to dynamically modulate two-qubit interactions.  The relative merits of these two schemes follow from their requirements; tuning the interactions requires control of individual qubits between the two sweet spots, while ``always-on'' interactions consume additional physical resources\cite{Zhou:2002jg,Benjamin:2003gh,Satoh:2015uo}.

\section{Summary}
\label{sec:sum}

In conclusion, we have studied an interface-bound silicon acceptor-spin qubit driven by a Rashba-like spin-orbit interaction induced by the interface. We have demonstrated that quadrupolar terms in the qubit interaction with the electric field are the source of sweet spots in the dephasing time, which enhance the qubit immunity to charge noise. The spin off-diagonal terms induced by the interface enhance EDSR significantly, which allows efficient manipulation entirely by means of electric fields. Entanglement by means of dipole-dipole interactions or cQED is inherently scalable thanks to the noise immunity of the qubit. The Rashba-like spin-orbit interaction terms improve the possibility of building a true spin qubit in solid state quantum computing platforms.

\acknowledgements

This work is supported by the ARC through the DP scheme. JS and SR acknowledge funding from the ARC Centre of Excellence for Quantum Computation and Communication Technology (CE110001027), and in part by the US Army Research Office (W911NF-08-1-0527). We thank S. Mahapatra, R.~Winkler, U.~Zuelicke, J.~Q.~You, W. Coish, S. Chesi, S. Ganguly, P. Raychaudhuri, R. Sensarma, P. Ayyub for useful discussions.

\appendix

\begin{widetext}

\section{Wave functions}
\label{ref:wavefun}
$\bm{S}$ can be regarded as a real spin-3/2 operator with eigenstates being spinors $\chi_{\pm\frac{3}{2}}$ and $\chi_{\pm\frac{1}{2}}$, while the eigenstates of $\bm{L}$ are the spherical harmonics $Y_{lm}$. For the convenience of evaluating the matrix elements of the interface potential, it is useful to decomposite the wave functions into uncoupled L-S representation using Clebsch-Gordan coefficients:

\begin{equation}
\begin{array}{rl}
\displaystyle \ket{\Psi_{\frac{3}{2}}} = & \displaystyle f_0(r) Y_{00} \, \chi_{\frac{3}{2}} + g_0(r) \bigg(\sqrt{\frac{2}{5}} \, Y_{22}\chi_{-\frac{1}{2}} - \sqrt{\frac{2}{5}} \, Y_{21}\chi_{\frac{1}{2}} + \sqrt{\frac{1}{5}} \, Y_{20}\chi_{\frac{3}{2}} \bigg) \\ [3ex]
\displaystyle \ket{\Psi_{\frac{1}{2}}} = & \displaystyle f_0(r) Y_{00} \, \chi_{\frac{1}{2}} + g_0(r) \bigg(\sqrt{\frac{2}{5}} \, Y_{22}\chi_{-\frac{3}{2}} - \sqrt{\frac{1}{5}} \, Y_{20}\chi_{\frac{1}{2}} + \sqrt{\frac{2}{5}} \, Y_{2,-1}\chi_{\frac{3}{2}} \bigg) \\ [3ex]
\displaystyle \ket{\Psi_{-\frac{1}{2}}} = & \displaystyle f_0(r) Y_{00} \, \chi_{-\frac{1}{2}} + g_0(r) \bigg(\sqrt{\frac{2}{5}} \, Y_{2,-2}\chi_{\frac{3}{2}} - \sqrt{\frac{1}{5}} \, Y_{20}\chi_{-\frac{1}{2}} + \sqrt{\frac{2}{5}} \, Y_{2,1}\chi_{-\frac{3}{2}} \bigg) \\ [3ex]
\displaystyle \ket{\Psi_{-\frac{3}{2}}} = & \displaystyle f_0(r) Y_{00} \, \chi_{-\frac{3}{2}} + g_0(r) \bigg( \sqrt{\frac{2}{5}} \, Y_{2,-2}\chi_{\frac{1}{2}} - \sqrt{\frac{2}{5}} \, Y_{2,-1}\chi_{-\frac{1}{2}} + \sqrt{\frac{1}{5}} \, Y_{20}\chi_{-\frac{3}{2}} \bigg).
\end{array}
\end{equation}
\begin{equation}
\begin{array}{rl}
\ket{\Phi_\frac{3}{2}}&=f_2(r)\bigg(-\sqrt{\dfrac{3}{5}}Y_{10}\chi_{\frac{3}{2}}+\sqrt{\dfrac{2}{5}}Y_{11}\chi_{\frac{1}{2}}\bigg)\\
&+g_2(r)\bigg(-\sqrt{\dfrac{1}{35}}Y_{30}\chi_{\frac{3}{2}}+\sqrt{\dfrac{4}{35}}Y_{31}\chi_{\frac{1}{2}}-\sqrt{\dfrac{2}{7}}Y_{32}\chi_{-\frac{1}{2}}+\sqrt{\dfrac{4}{7}}Y_{33}\chi_{-\frac{3}{2}}\bigg)\\
\ket{\Phi_\frac{1}{2}}&=f_2(r)\bigg(-\sqrt{\dfrac{2}{5}}Y_{1,-1}\chi_{\frac{3}{2}}-\sqrt{\dfrac{1}{15}}Y_{10}\chi_{\frac{1}{2}}+2\sqrt{\dfrac{2}{15}}Y_{11}\chi_{-\frac{1}{2}}\bigg)\\
&+g_2(r)\bigg(-\sqrt{\dfrac{4}{35}}Y_{3,-1}\chi_{\frac{3}{2}}+\sqrt{\dfrac{9}{35}}Y_{30}\chi_{\frac{1}{2}}-2\sqrt{\dfrac{3}{35}}Y_{31}\chi_{-\frac{1}{2}}+\sqrt{\dfrac{2}{7}}Y_{32}\chi_{-\frac{3}{2}}\bigg)\\
\ket{\Phi_{-\frac{1}{2}}}&=f_2(r)\bigg(\sqrt{\dfrac{2}{5}}Y_{1,1}\chi_{-\frac{3}{2}}+\sqrt{\dfrac{1}{15}}Y_{10}\chi_{-\frac{1}{2}}-2\sqrt{\dfrac{2}{15}}Y_{1,-1}\chi_{\frac{1}{2}}\bigg)\\
&+g_2(r)\bigg(\sqrt{\dfrac{4}{35}}Y_{31}\chi_{-\frac{3}{2}}-\sqrt{\dfrac{9}{35}}Y_{30}\chi_{-\frac{1}{2}}+2\sqrt{\dfrac{3}{35}}Y_{3,-1}\chi_{\frac{1}{2}}-\sqrt{\dfrac{2}{7}}Y_{3,-2}\chi_{\frac{3}{2}}\bigg)\\
\ket{\Phi_{-\frac{3}{2}}}&=f_2(r)\bigg(\sqrt{\dfrac{3}{5}}Y_{10}\chi_{-\frac{3}{2}}-\sqrt{\dfrac{2}{5}}Y_{1,-1}\chi_{-\frac{1}{2}}\bigg)\\
&+g_2(r)\bigg(\sqrt{\dfrac{1}{35}}Y_{30}\chi_{-\frac{3}{2}}-\sqrt{\dfrac{4}{35}}Y_{3,-1}\chi_{-\frac{1}{2}}+\sqrt{\dfrac{2}{7}}Y_{3,-2}\chi_{\frac{1}{2}}-\sqrt{\dfrac{4}{7}}Y_{3,-3}\chi_{\frac{3}{2}}\bigg)
\end{array}
\end{equation}

\section{Eigenvector matrix and eigenvalues}
\label{app:eig}
The eigenvector matrix $V$ in the basis $\{\ket{v_{H+}},\ket{v_{H-}},\ket{v_{L+}},\ket{v_{L-}}\}$ takes the form
\begin{equation}
\arraycolsep0.3ex
\begin{array}{rl}
\displaystyle V = & \displaystyle \begin{pmatrix}
-ia/N_1 & 0 & 0 & 1/N_1 \cr
0 & ib/N_2 & 1/N_2 & 0 \cr
0 & 1/N_2 & ib/N_2 & 0 \cr
1/N_1 & 0 & 0 & -ia/N_1
\end{pmatrix} \\ [3ex]
\displaystyle V^\dag = & \displaystyle \begin{pmatrix}
ia/N_1 & 0 & 0 & 1/N_1 \cr 
0 & -ib/N_2 & 1/N_2 & 0 \cr
0 & 1/N_2 & -ib/N_2 & 0 \cr
1/N_1 & 0 & 0 & ia/N_1
\end{pmatrix}.
\end{array}
\label{eigenvectors}
\end{equation}
where $N_1,N_2,a,b$ are normalisation factors introduced for typographical simplicity, and are given by:
\begin{equation}
\begin{array}{c}
a=-\dfrac{2\sqrt{3}pF_z}{(\varepsilon_{H+}-\varepsilon_{L-})+\sqrt{(\varepsilon_{H+}-\varepsilon_{L-})^2+12p^2F_z^2}}\\
b=-\dfrac{2\sqrt{3}pF_z}{(\varepsilon_{H-}-\varepsilon_{L+})+\sqrt{(\varepsilon_{H-}-\varepsilon_{L+})^2+12p^2F_z^2}}\\
N_1=\sqrt{1+a^2}\\
N_2=\sqrt{1+b^2}
\end{array}
\end{equation}

For $\varepsilon_Z<\frac{1}{2}\Delta_{HL}$, the eigenvalues in ascending order are
\begin{equation}
\begin{array}{rl}
\lambda_{H-}&=\dfrac{1}{2}[(\varepsilon_{H-}+\varepsilon_{L+})-\sqrt{(\varepsilon_{H-}-\varepsilon_{L+})^2+12p^2F_z^2}]\\
\lambda_{H+}&=\dfrac{1}{2}[(\varepsilon_{H+}+\varepsilon_{L-})-\sqrt{(\varepsilon_{H+}-\varepsilon_{L-})^2+12p^2F_z^2}]\\
\lambda_{L-}&=\dfrac{1}{2}[(\varepsilon_{H+}+\varepsilon_{L-})+\sqrt{(\varepsilon_{H+}-\varepsilon_{L-})^2+12p^2F_z^2}]\\
\lambda_{L+}&=\dfrac{1}{2}[(\varepsilon_{H-}+\varepsilon_{L+})+\sqrt{(\varepsilon_{H-}-\varepsilon_{L+})^2+12p^2F_z^2}]
\end{array}
\label{SmallZeemanEigenvalue}
\end{equation}
For $\varepsilon_Z>\frac{1}{2}\Delta_{HL}$, the eigenvalues in ascending order are:
\begin{equation}
\begin{array}{rl}
\lambda_{H-}&=\dfrac{1}{2}[(\varepsilon_{H-}+\varepsilon_{L+})-\sqrt{(\varepsilon_{H-}-\varepsilon_{L+})^2+12p^2F_z^2}]\\
\lambda_{L-}&=\dfrac{1}{2}[(\varepsilon_{H+}+\varepsilon_{L-})-\sqrt{(\varepsilon_{H+}-\varepsilon_{L-})^2+12p^2F_z^2}]\\
\lambda_{H+}&=\dfrac{1}{2}[(\varepsilon_{H+}+\varepsilon_{L-})+\sqrt{(\varepsilon_{H+}-\varepsilon_{L-})^2+12p^2F_z^2}]\\
\lambda_{L+}&=\dfrac{1}{2}[(\varepsilon_{H-}+\varepsilon_{L+})+\sqrt{(\varepsilon_{H-}-\varepsilon_{L+})^2+12p^2F_z^2}]
\end{array}
\label{LargeZeemanEigenvalue}
\end{equation}
This can be understood by the fact that $\lambda_{H\pm}$ and $\lambda_{L\pm}$ tend to $\varepsilon_{H\pm}$ and $\varepsilon_{L\pm}$ in the limit $F_z=0$. For the construction of a qubit the two states of which have opposite spin orientation, we should guarantee that $\varepsilon_Z<\frac{1}{2}\Delta_{HL}$.

For the discussion of relaxation, we further define $\lambda_{H/L}$ to be the eigenvalues in \ref{SmallZeemanEigenvalue} without Zeeman field:
\begin{equation}
\begin{array}{rl}
\lambda_H=&\lambda_{H\pm}\big |_{\varepsilon_Z=0}=\dfrac{1}{2}[(\varepsilon_H+\varepsilon_L)-\sqrt{(\varepsilon_H-\varepsilon_L)^2+12p^2F_z^2}]\\
\lambda_L=&\lambda_{L\pm}\big |_{\varepsilon_Z=0}=\dfrac{1}{2}[(\varepsilon_H+\varepsilon_L)+\sqrt{(\varepsilon_H-\varepsilon_L)^2+12p^2F_z^2}]
\end{array}
\label{EigNoZeeman}
\end{equation}
Then the splitting between heavy hole subspace and light hole subspace becomes 
\begin{equation}
\tilde{\Delta}_{HL}=\lambda_L-\lambda_H=\sqrt{\Delta_{HL}^2+12p^2F_z^2}
\label{Splitting}
\end{equation}

\section{Interface spin-orbit Hamiltonian after the Schrieffer-Wolff transformation}
\label{app:SW}

We work out $H_{if}+H_{gt}=U_0\Theta(-z)+ e\bm E\cdot\bm r$ in the basis $\{\Psi_{3/2},\Psi_{1/2},\Psi_{-1/2},\Psi_{-3/2},\Phi_{3/2},\Phi_{1/2},\Phi_{-1/2},\Phi_{-3/2}\}$ using the wave functions calculated in Appendix \ref{ref:wavefun}. Projecting all terms that couple $\ket{\Phi}$ states and $\ket{\Psi}$ states into the ground state manifold using Schrieffer-Wolff transformation, keeping terms up to the second order, we obtain the effective Rashba Hamiltonian in basis $\{3/2,-3/2,1/2,-1/2\}$:

\begin{equation}
\arraycolsep0.3ex
\begin{array}{rl}
\displaystyle H_{if}+H_{gt} = & \displaystyle -\frac{1}{\Delta} \, \begin{pmatrix}
(U_H +\sqrt{3} e\rho F_z)^2 & 0& e E_- \rho (U_t +\dfrac{4\sqrt{3}}{3} e \rho F_z) & 0 \cr  
0 & (U_H + \sqrt{3} e\rho F_z)^2 & 0 & - e E_+ \rho (U_t + \dfrac{4\sqrt{3}}{3} e \rho F_z) \cr
e E_+ \rho (U_t + \dfrac{4\sqrt{3}}{3} e \rho F_z)& 0 & (U_L + \dfrac{\sqrt{3}}{3} e \rho F_z)^2 & 0 \cr
0 & - e E_- \rho (U_t + \dfrac{4\sqrt{3}}{3} e \rho F_z)& 0 & (U_L + \dfrac{\sqrt{3}}{3} e \rho F_z)^2
\end{pmatrix}
\end{array}
\end{equation}

where $\Delta=\varepsilon_1-\varepsilon_0$ is the splitting between ground state and the first excited state,
$\rho=-\sqrt{\dfrac{1}{15}}\tbkt{f_0}{r}{f_2}-\dfrac{4}{5}\sqrt{\dfrac{1}{15}}\tbkt{g_0}{r}{f_2}-\dfrac{3}{5}\sqrt{\dfrac{1}{15}}\tbkt{g_0}{r}{g_2}$ comes from the radial integral, 
$U_H=\tbkt{\Psi_{\frac{3}{2}}}{H_{if}}{\Phi_{\frac{3}{2}}}$, 
$U_L=\tbkt{\Psi_{\frac{1}{2}}}{H_{if}}{\Phi_{\frac{1}{2}}}$, 
$U_t=U_H+U_L$ all come from the interface.

The off-diagonal terms are to couple between heavy holes and light holes which can be simplified as:

\begin{equation}
\arraycolsep0.3ex
\begin{array}{rl}
\displaystyle H_R = & \displaystyle 
(a_R + b_R F_z) \, \begin{pmatrix}
0 & 0& E_- & 0 \cr  
0 & 0 & 0 & - E_+ \cr
E_+ & 0 & 0 & 0 \cr
0 & - E_- & 0 & 0
\end{pmatrix}.
\end{array}
\label{off-diagonal}
\end{equation}
where $a_R=-\dfrac{e\rho U_t }{\Delta}$ comes from the interface, while $b_R=-\dfrac{4\sqrt{3}e^2\rho^2}{3\Delta}$ describes the (first-order) coupling due to the gate field $F_z$.

The diagonal terms give a parabolic correction to the splitting between heavy holes and light holes:
\begin{equation}
\begin{array}{c}
\varepsilon_H\rightarrow\varepsilon_0+u_H-\dfrac{(U_H+\sqrt{3} e\rho F_z)^2}{\Delta}\\
\varepsilon_L\rightarrow\varepsilon_0+u_L-\dfrac{(3U_L+\sqrt{3} e \rho F_z)^2}{9\Delta}\\
\Delta_{HL} \approx \varepsilon_L-\varepsilon_H=\Delta_{HL}^{(0)}+\Delta_{HL}^{(1)}F_z+\Delta_{HL}^{(2)}F_z^2
\end{array}
\label{RashbaSplitting}
\end{equation}
where $\Delta_{HL}^{(0)}=u_L-u_H-\dfrac{1}{\Delta}(U_L^2-U_H^2)$, $\Delta_{HL}^{(1)}=\dfrac{2\sqrt{3} e^2\rho}{3\Delta}(U_L-3 U_H)$, $\Delta_{HL}^{(2)}=\dfrac{8e^2\rho^2}{3\Delta}$ in this case.

\section{Relaxation}
\label{app:rel}

We derive the phonon-induced spin relaxation for a $J=3/2$ system.  In first-order perturbation theory, the phonon-mediated relaxation rate from a level $\left|n'\right\rangle$ to a level $\left|n\right\rangle$ via emission of a phonon with energy $\hbar\omega_{qs}=\hbar v_s q_s$ is
\begin{equation}
\frac{1}{T_{n\rightarrow n'}}=\frac{2\pi}{\hbar}\sum_{i,j,s,\mathbf{q}_s}|\left\langle n',n_{q}+1|H_{\epsilon ij s}|n,n_q\right\rangle|^2\delta(\varepsilon_n-\varepsilon_n'-\hbar\omega_{qs})
\end{equation}
where $s=\ell,t_1,t_2$ are the phonon polarizations, $\mathbf{q}_s$ is the phonon wavevector, and $\sum_{i,j}H_{\epsilon ij s}=\sum_{i,j}D_{ij}\epsilon_{ijs}$. The deformation potential matrices $D_{ij}$ are determined from the Bir-Pikus Hamiltonian
\begin{align}
\sum D_{ij}\epsilon_{ijs}=&a'(\epsilon_{xxs}+\epsilon_{yys}+\epsilon_{zzs})\nonumber\\+&b'[(J_x^2-\tfrac{5}{4}I)\epsilon_{xxs}+(J_y^2-\tfrac{5}{4}I)\epsilon_{yys}+(J_z^2-\tfrac{5}{4}I)\epsilon_{zzs}]\nonumber\\+&(2d'/\sqrt{3})[\{J_x,J_y\}\epsilon_{xys}+\{J_y,J_z\}\epsilon_{yzs}+\{J_x,J_z\}\epsilon_{xzs}]
\end{align} 
and the strain $\epsilon_{ijs}=\tfrac{1}{2}(d\delta R_{is}/dr_j+d\delta R_{js}/dr_i)$ of the phonon mode $s$ is determined by the displacement \cite{Srivastava}
\begin{equation}
\delta\mathbf{R}_s=(-i)\sqrt{\frac{\hbar}{2NV_c\rho\omega_{qs}}}\hat{\mathbf{e}}_{\mathbf{q}s}(a^\dagger_{\mathbf{q}s}+a_{\mathbf{q}s})\exp(i\mathbf{q}_s\cdot\mathbf{r}),
\end{equation}
where $\hat{\mathbf{e}}_{qs}$ is the normalized phonon polarization vector, $N$ is the number of unit cells, $V_c$ is the unit cell volume, $NV_c=L^3$ is the crystal volume, and $\rho$ is the mass density.  Using these relations, we obtain
\begin{equation}
H_{\epsilon i j s}=q D_{ij}\sqrt{\frac{\hbar}{2NV_c\rho\omega_{qs}}}\Big(-\frac{i}{2}\Big)(e_{qis}q_{j}/q+e_{qjs}q_{i}/q)(a^\dagger_{\mathbf{q}s}+a_{\mathbf{q}s})\exp(i\mathbf{q}_s\cdot\mathbf{r})
\end{equation}
For $\left\langle n'|D_{ij}\exp(i\mathbf{q}\cdot\mathbf{r})|n\right\rangle$ we make use of the dipole approximation $\left\langle n'|D_{ij}(1+i\mathbf{q}\cdot\mathbf{r}+\dots)|n\right\rangle\approx\left\langle n'|D_{ij}|n\right\rangle$, which is appropriate since $qa\sim10^{-2}$ where $q$ is the the phonon wavevector for $\hbar\omega=60$ $\mu$eV and $a\sim1$ nm is the Bohr radius.   The identity $\delta[g(x)]=\sum_i\delta(x-x_i)/|(\partial g/\partial x)_{x=x_i}|$, where $x_i$ are the zeros of $g(x)$, gives $\delta(\hbar\omega-\hbar v_q q_{s})=\delta(q-\omega/v_q)/(\hbar v_q)$, and we furthermore the summation over phonon wavevectors to an integral with $\sum_q(\cdot)\approx (\frac{L}{2\pi})^3\int dq^3 (\cdot)$.  Using these simplifications we get
\begin{equation}
\frac{1}{T_{n\rightarrow n'}}=\frac{1}{32\hbar\pi^2\rho}\sum_{i,j,s}|\left\langle n' | D_{ij}| n \right\rangle|^2\frac{\Omega_{ijs}}{v_{qs}^2}\int q^3 dq \delta(q-\omega/v_q)|\left\langle n_{q}+1|(a^\dagger_{\mathbf{q}s}+a_{\mathbf{q}s})|n_q\right\rangle|^2 
\end{equation}
where $\Omega_{ijs}=\int d\Omega (e_{qs i}q_j/q+e_{qs j}q_i/q)^2$ is a dimensionless angular integral.  At low temperatures $T_\omega\ll \hbar\omega/k \approx 0.7$ K we have $|\left\langle n_{q}+1|(a^\dagger_{\mathbf{q}s}+a_{\mathbf{q}s})|n_q\right\rangle|^2\approx 1$, and
\begin{equation}
\frac{1}{T_{n\rightarrow n'}}=\frac{1}{32\hbar^4\pi^2\rho}\sum_{i,j}|\left\langle n' | D_{ij}| n \right\rangle|^2\Big(\frac{\Omega_{ij\ell}}{v_{l}^5}+\frac{\Omega_{ijt_1}+\Omega_{ijt2}}{v_{t}^5}\Big)
\end{equation}
where $v_l$ and $v_t$ are the longitudinal and transverse acoustic phonon velocities. The $\Omega_{ijs}$ are computed by substitution of $\mathbf{q}_s$ and $\hat{e}_{qs}$ \cite{Ehrenreich_PRB1956}
\begin{align}
\hat{e}_{q\ell}&=q^{-1}(q_x,q_y,q_z)\\
&=(\cos(\theta)\sin(\phi),\sin(\theta)\sin(\phi),\cos(\phi))\\
\hat{e}_{qt_1}&=(q_x^2+q_y^2)^{-\tfrac{1}{2}}(q_y,-q_x,0)\\
&=(\sin(\theta),-\cos(\theta),0)\\
\hat{e}_{qt_2}&=q^{-1}(q_x^2+q_y^2)^{-\tfrac{1}{2}}(q_xq_z,q_yq_z,-(q_x^2+q_y^2))\\
&=(\cos(\theta)\cos(\phi),\sin(\theta)\cos(\phi),-\sin(\phi)).
\end{align}
For the longitudinal phonons we have $\Omega_{xx\ell}=\Omega_{yy\ell}=\Omega_{zz\ell}=16\pi/5$ and $\Omega_{xy\ell}=\Omega_{yz\ell}=\Omega_{xz\ell}=16\pi/15$.  Similarly, for the ta1 phonons we have $\Omega_{xxt1}=\Omega_{yyt1}=4\pi/3$ and $\Omega_{zzt1}=0$, and $\Omega_{xyt1}=4\pi/3$ and $\Omega_{yzt2}=\Omega_{xzt2}=2\pi/3$. Finally, for the ta2 phonons we have $\Omega_{xxt2}=\Omega_{yyt2}=4\pi/5$ and $\Omega_{zzt2}=32\pi/15$ and $\Omega_{xyt2}=4\pi/15$ and $\Omega_{yzt2}=\Omega_{xzt2}=14\pi/15$. These integrals obey $\Omega_{ij\ell}=16\pi/5$ for $i=j$, $\Omega_{ij\ell}=16\pi/15$ for $i\ne j$, $\Omega_{ijt1}+\Omega_{ijt2}=32\pi/15$ for $i=j$ and $\Omega_{ijt1}+\Omega_{ijt2}=8\pi/5$ for $i\ne j$. 
Substituting these expressions we obtain
\begin{equation}
\frac{1}{T_1}=\frac{(\hbar\omega)^3}{20\hbar^4\pi\rho}\Big[\sum_i |\left\langle n' | D_{ii}| n \right\rangle|^2 \Big(\frac{2}{v_l^5}+\frac{4}{3v_t^5}\Big)+\sum_{i\neq j}|\left\langle n' | D_{ij}| n \right\rangle|^2\Big(\frac{2}{3v_l^5}+\frac{1}{v_t^5}\Big)\Big]
\end{equation}

For the heavy hole qubit spin relaxation is given by
\begin{equation}
\frac{1}{T_1}=\frac{(\hbar\omega)^3}{20\hbar^4\pi\rho}\Big[\sum_i |\left\langle - | {D}_{ii}| + \right\rangle|^2 \Big(\frac{2}{v_l^5}+\frac{4}{3v_t^5}\Big)+\sum_{i\neq j}|\left\langle - | {D}_{ij}| + \right\rangle|^2\Big(\frac{2}{3v_l^5}+\frac{1}{v_t^5}\Big)\Big]
\end{equation}

We determine the deformation potential $\left\langle - | D_{ij}| + \right\rangle$ of the heavy-hole spin-orbit qubit using a Schrieffer-Wolff transformation.  Because the magnetic field is a diagonal interaction for the heavy-hole qubit, we work with a perturbation  $\bar{H'}=\bar{H}_{Z}+\bar{H}_{\rm hp}$, where $\bar{H}_{\rm hp}=V^\dagger(\sum_{i,j,s} D_{ij}\epsilon_{ijs})V$, and $\bar{H}_{Z}=V^\dagger H_Z V$. 
\begin{equation}
\left\langle - | D_{ij}| + \right\rangle\approx\frac{1}{\tilde{\Delta}_{HL}}(\tilde{H}'_{-,L-}\tilde{H}'_{L-,+}+\tilde{H}'_{-,L+}\tilde{H}'_{L+,+}).
\end{equation}
where $\tilde{H}'=\tilde{H}_{Zo}+\tilde{D}_{ij}$.  For the HH qubit we obtain $|\left\langle - | D_{xx}| + \right\rangle|^2=0$, $|\left\langle - | D_{yy}| + \right\rangle|^2=0$,
$|\left\langle - | D_{zz}| + \right\rangle|^2=0$,$|\left\langle - | D_{xy}| + \right\rangle|^2=0$, $|\left\langle - | D_{yz}| + \right\rangle|^2=3d'^2(\varepsilon_Z/\tilde{\Delta}_{HL})^2$, and $|\left\langle - | D_{xz}| + \right\rangle|^2=3d'^2(\varepsilon_Z/\tilde{\Delta}_{HL})^2$.  Collecting terms, we have
\begin{equation}
\frac{1}{T_1}=\frac{(\hbar\omega)^3}{20\hbar^4\pi\rho}\Big(\frac{\varepsilon_Z}{\tilde{\Delta}_{HL}}\Big)^2\Big[6d'^2\Big(\frac{2}{3v_l^5}+\frac{1}{v_t^5}\Big)\Big]
\end{equation}
where $\varepsilon_Z=g\mu_BB$ and $\hbar\omega$ is the qubit frequency.  This value of $T_1^{-1}$ is approximately $3(\varepsilon_Z/\tilde{\Delta}_{HL})^2$ smaller than the value for the $\left|-1/2\right\rangle$ to $\left|-3/2\right\rangle$ transition in Si:B.  The factor of 3 comes because the Zeeman energy is tripled, and the factor $(\varepsilon_Z/\tilde{\Delta}_{HL})^2$ comes because the transition here is for a spin qubit. We use $v_l=8.99\times10^{3}$ m/s and $v_t=1.7v_l$ for the longitudinal and transverse sound velocities in silicon, respectively, $\rho=2330$ kg/m$^3$ for the mass density, and $b'=-1.4$ eV and $d'=-3.7$ eV for the Bir-Pikus deformation potentials.  We obtain $T_1=40$ $\mu$s for $B=0.5$ T, and $T_1=1.3$ ms for $B=0.25$ T. 

\begin{figure}
\includegraphics[width=\columnwidth]{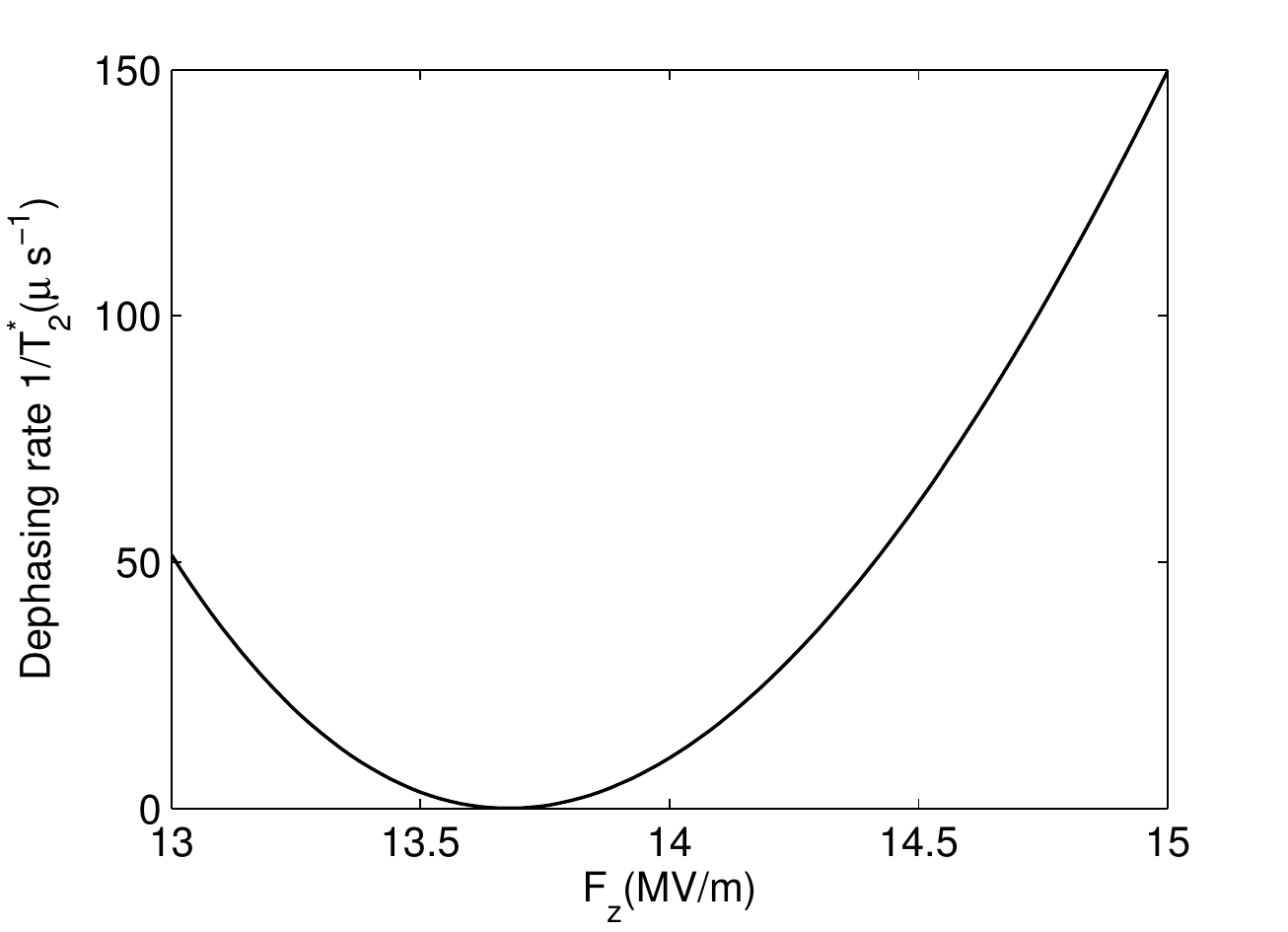}
\caption{Dependence of $T_2^*$ on gate electric field for $z_0=4.6$ nm around the sweet spot.}
\label{HeavyDephasingInset}
\end{figure}

\section{Dephasing}
\label{app:deph}

The in-plane components of the defect electric field do not contribute to dephasing in leading order. The eigenvalues of the two lowest qubit states $\lambda_{H\pm}$ are expanded in $\mathcal{F}_z$ as $\lambda_{H\pm} = \lambda^{(0)} + C_{H\pm} \mathcal{F}_z$, which can be written as
\begin{equation}
\lambda_{H\pm}=\lambda^{(0)}_{H\pm} +\dfrac{C_{H-}+C_{H+}}{2} \, \mathcal{F}_z + \dfrac{C_{H-}-C_{H+}}{2} \, \mathcal{F}_z \, \sigma_z.
\end{equation}
The dephasing rate \cite{Bermeister_APL14} $1/T_2^* = V^2\tau/(2\hbar^2)$ where the energy fluctuation due to the defect field is given by $V = (C_{H-}-C_{H+}) \mathcal{F}_z/2$. Using Eq.(\ref{SmallZeemanEigenvalue}), 
\begin{equation}
\begin{array}{rl}
\displaystyle \dfrac{1}{T_2^*} = &\dfrac{\mathcal{F}_z\tau}{32\hbar^2}\bigg[\dfrac{(\Delta_{HL}^{(1)}+2\Delta_{HL}^{(2)}F_z)(\Delta_{HL}-2\varepsilon_Z)+12p^2F_z^2}{\sqrt{(\Delta_{HL}-2\varepsilon_Z)^2+12p^2F_z^2}}\\
& - \dfrac{(\Delta_{HL}^{(1)}+2\Delta_{HL}^{(2)}F_z)(\Delta_{HL}+2\varepsilon_Z)+12p^2F_z^2}{\sqrt{(\Delta_{HL}+2\varepsilon_Z)^2+12p^2F_z^2}}\bigg]^2.
\end{array}
\end{equation}

\end{widetext}


\end{document}